\documentclass[11pt]{article}
\usepackage{amsmath,amssymb,color,graphics,epsfig,cite,subfloat,amsthm,graphicx,caption,
epstopdf,float,subfig,subfloat,geometry,blindtext,indentfirst,cite,hyperref,cleveref}

\usepackage{amsfonts}

\newcommand{\be}{\begin{equation}}
\newcommand{\ee}{\end{equation}}
\newcommand{\bea}{\setlength\arraycolsep{2pt} \begin{eqnarray}}
\newcommand{\eea}{\end{eqnarray}}

\def\fft#1#2{{\frac{#1}{#2}}}

\def\0{{\sst{(0)}}}
\def\1{{\sst{(1)}}}
\def\2{{\sst{(2)}}}
\def\3{{\sst{(3)}}}
\def\4{{\sst{(4)}}}
\def\5{{\sst{(5)}}}
\def\6{{\sst{(6)}}}
\def\7{{\sst{(7)}}}
\def\8{{\sst{(8)}}}
\def\sst#1{{\scriptscriptstyle #1}}

\begin{document}

\begin{center}
{\Large {\bf Scalarization of Taub-NUT Black Holes in Extended scalar-tensor-Gauss-Bonnet Theory }}

\vspace{20pt}

{\large Hai-Shan Liu and Lei Zhang}

\vspace{10pt}

\vspace{10pt}

{\it Center for Joint Quantum Studies and Department of Physics,\\
School of Science, Tianjin University, Tianjin 300350, China }

\vspace{40pt}

\underline{ABSTRACT}

\end{center}

Recently, scalarization of Schwarzschild black hole are extensively studied. In this work, we explore the scalarization of Taub-NUT black hole. The theory we consider is the extended scalar-tensor-Gauss-Bonnet theory, which admits Ricci-flat Taub-NUT black hole as a solution. An analysis of probe scalar field is carried out to identify the mass parameter and NUT parameter $(m\,,n)$ where the hairy black holes start to emerge. Then, we use shooting method to construct the scalarized Taub-NUT black hole numerically. Being different from the Schwarzschild case, there exists two branches of new hairy black holes which are smoothly connected to each other. We calculate the entropy of scalarized black holes and compare it with the entropy of scalar-free Taub-NUT black holes, it turns out that the entropy of the new hairy black holes are larger than that of scalar-free black holes. A novel phenomena emerges in this system that the entropy of the black holes at the bifurcation point is constant for positive mass parameter. We then conjecture a maximal entropy bound for all the scalarized black hole whose mass parameter at the bifurcation point is greater than zero.

\vfill{\footnotesize  hsliu.zju@gmail.com ~~~~ leiz@tju.edu.cn  }


\thispagestyle{empty}
\pagebreak



\section{Introduction}
Whether there exists black holes carrying scalar hair has been attracting much attentions for years. And a no-hair theorem was proposed, which states that Einstein's general relativity can not support static black holes with a non-trivial scalar\cite{Bekenstein:1972ny,Bekenstein:1995un}. Recently, there is a big breakthrough in this direction. Many black holes with non-trivial scalar field have been successfully constructed in the extended scalar-tensor theories\cite{Doneva:2017bvd,Antoniou:2017hxj,Silva:2017uqg,Minamitsuji:2018xde,Cunha:2019dwb}. These black holes are resulting from the scalarization procedure and are often called scalarized black holes, more details can be found in\cite{Damour:1993hw,Myung:2018vug}. It is found that a wide class of extended scalar-tensor theories can allow spontaneous scalarization, which means that the theories can accommodate scalarized black holes as well as the standard vacuum solutions of general relativity.

Schwarzschild black hole is the simplest black hole solutions of general relativity, and the scalarization processes which are connected to Schwarzschild black hole have been  extensively studied. It is well-known that Taub-NUT spacetime is also a solution of general relativity\cite{Taub:1950ez,Newman:1963yy}. The Taub-NUT solution has two integration constants, the mass parameter and NUT parameter, when the NUT parameter approaches zero, the solution returns back to Schwarzschild black hole solution. Though the Taub-NUT black hole solution was discovered in the middle of last century, the  thermodynamics of this solution is still controversial, especially, the definition and physical interpretation of its mass and NUT charge.  There are some new progress on this subject in the last few years \cite{NUTthmann1,NUTth2,NUTth3}, triggering a new surge of interest in this area. In this work, we shall study the scalarization of black holes which are connected to Taub-NUT black holes, providing a wider platform to explore the thermodynamics of Taub-NUT like black holes.

We try to obtain Taub-NUT black holes with non-trivial scalar fields through coupling the scalar field directly to spacetime curvature invariants. In particular, we shall consider the extended scalar-tensor-Gauss-Bonnet theories in four dimensions.  It is shown that this theory can support spontaneous scalarization of Schwarzschild black holes for certain classes of coupling functions, the Schwarzschild black hole will become unstable below a critical value of mass, then new scalarized black holes will emerge. We shall choose one coupling function such that the theory can admit Taub-NUT black hole as a solution. A probe scalar field analysis will be carried out to identify the bifurcation points where scalarized black holes start to appear. Then we shall apply shooting method to construct the scalarized Taub-NUT black holes numerically.

The structure of this paper is as follows. In section 2, we present the extended scalar-tensor-Gauss-Bonnet gravity theory with a particular coupling function which admit Taub-NUT black hole as a solution and take a probe scalar analysis to identify parameter range where the scalarized black holes start emerge. In section 3, we analysis the behavior of Taub-NUT like black holes near the horizon and in the infinity, respectively. We numerically construct scalarized Taub-NUT black holes and study their properties in section 4. Finally, a conclusion of our results are present in Section 5.

\section{Extended scalar-tensor-Gauss-Bonnet theory}
In this section, we consider extended scalar-tensor-Gauss-Bonnet theory, which is given by
\begin{equation}
		S=\frac{1}{16\pi}\int d^4x\sqrt{-g}\left(R-2\nabla_{\mu}\varphi\nabla^{\mu}\varphi-V\left(\varphi\right)+\lambda ^2f\left(\varphi\right){R_{GB}}^2\right) \,,
\end{equation}

where $R$ is the Ricci scalar with respect of the metric $g_{\mu\nu}$, $\varphi$ is the scalar field with the potential $V(\varphi)$,  $f(\varphi)$ is  the coupling function with coupling constant $\lambda$  and $R_{GB}^2$ is the Gauss-Bonnet invariant.

\begin{equation}
	R_{GB}^2=R^2+R_{\mu\nu\rho\lambda}R^{\mu\nu\rho\lambda}-4R_{\mu\nu}R^{\mu\nu}
\end{equation}
	
	The equations of motion of the theory can be obtained  from the variation of the action(1) with respect to the metric $g_{\mu\nu}$ and the scalar field $\varphi$,
	\begin{equation}
		R_{\mu\nu}-\frac{1}{2}Rg_{\mu\nu}+\varGamma_{\mu\nu}=2\nabla_{\mu}\varphi\nabla_{\nu}\varphi-g_{\mu
			\nu}\nabla_{\alpha}\varphi\nabla^{\alpha}\varphi-\frac{1}{2}g_{\mu\nu}V\left(\varphi\right) \,,
	\end{equation}
	\begin{equation}
		\nabla_{\alpha}\nabla^{\alpha}\varphi=\frac{1}{4}\frac{dV(\varphi)}{d\varphi}-\frac{\lambda^2}{4}\frac{dF(\varphi)}{d\varphi}R_{GB}^2
	\,.\end{equation}
	Here $\nabla_{\mu}$ is the covariant derivative with respect to the metric $g_{\mu\nu}$ and the $\varGamma_{\mu\nu}$ are defined by
	
	\begin{equation}
		\begin{split}
			\varGamma_{\mu\nu}=&-2R\nabla_{\mu}\varPsi_{\nu}-4\nabla^{\alpha}\varPsi_{\alpha}(R_{\mu\nu}-\frac{1}{2}Rg_{\mu\nu})+8R_{\mu\alpha}\nabla^{\alpha}\varPsi_{\nu}\\
			&-4g_{\mu\nu}R^{\alpha\beta}\nabla_{\alpha}\varPsi_{\beta}+4R^{\beta}_{\mu\alpha\nu}\nabla^{\alpha}\varPsi_{\beta}
		\,.\end{split}
	\end{equation}
	with $\varPsi_{\mu}$ given by
	
	\begin{equation}
		\varPsi_{\mu}=\lambda^2\frac{dF(\varphi)}{d\varphi}\nabla_{\mu}\varphi  \,.
	\end{equation}
	For simplicity,  we set the scalar potential $V(\varphi)$ to zero. It was shown that the theory can admit Schwarzschild black hole solution, when the coupling function satisfies $\fft{\delta f}{\delta \varphi}|_{\varphi = 0} = 0$\cite{Doneva:2017bvd}. Further, if   $\fft{\delta^2 f}{\delta \varphi ^2}|_{\varphi = 0} > 0$, scalarized black holes can emerge. Various coupling functions are studied for scalarization of Schwarzschild black holes\cite{Doneva:2017bvd,Antoniou:2017hxj} In this article, we choose the coupling function as that in \cite{Doneva:2017bvd}
		\begin{equation}
		F\left(\varphi\right)=\frac{1}{12}\left[1-\exp(-6\varphi^2)\right] \,.
	\end{equation}
It is easy to check that the above two conditions related to coupling function $F$  are satisfied. If the scalar field is trivial  $\varphi =0$, the theory will return to general relativity, and the equations of motion admit Ricci-flat Schwarzschild black hole as a solution.

As we mentioned in the introduction, the Ricci-flat Taub-NUT black hole solution is also a solution of general relativity. Thus,  Taub-NUT black hole with trivial scalar $\varphi=0$ is also a solution for the extended scalar-tensor-Gauss-Bonnet theory.
   The Taub-NUT metric has the form of\cite{Liu:2022wku}
    \begin{equation}
		ds^{2}=-f\left(dt+2n\cos\theta d\phi\right)^{2}+\frac{dr^2}{h}+\left(r^2+n^2\right)\left(d\theta^2+\sin^2\theta d\phi^2\right) \,
	\end{equation}
with
\be
h=f=\frac{r^2-2mr-n^2}{r^2+n^2} \,.
\ee
The Taub-NUT solution has two integration constants, one is the mass parameter $m$ and the other is $n$ which is usually called NUT parameter. As one can see, when $n=0$ the Taub-NUT solution goes back to Schwarzschild solution. In this article, we want to explore  scalarization of black hole under Taub-NUT background.

Following the strategy in \cite{Brihaye:2018bgc}, we first consider a test scalar field $\delta \varphi$, and the scalar equation on Taub-NUT background is given by
\begin{equation}
\nabla_{\alpha}\nabla^{\alpha}\delta\varphi+\frac{\lambda^2}{4}R_{GB}^2\delta\varphi=0 \,,
\end{equation}
the Gauss-Bonnet combination $R_{GB}^2$ on the Taub-NUT background is
\begin{align}
		R_{GB}^2=\frac{48}{\left(n^2+r^2\right)^6}&\left[m^2 \left(-n^6+15 n^4 r^2-15 n^2 r^4+r^6\right)+4 m \left(3 n^6 r-10 n^4 r^3+3 n^2 r^5\right)\notag\right.
		\\
		\phantom{=\;\;}
		&\left.+n^8-15 n^6 r^2+15 n^4 r^4-n^2 r^6\right]\,.
\end{align}
	
Then, it becomes an eigenvalue problem, under appropriate boundary conditions, fixing $(n\,, \lambda)$ the equation can admit solutions for a discrete mass parameter $m$. These points are the bifurcation points where the scalarized black hole start to emerge.
  Specifically, performing a spherical harmonics decomposition of the scalar field\cite{Herdeiro:2018wub} $\delta\varphi\left(r,\theta,\phi\right)=\sum_{lm}Y_{lm}\left(\theta,\phi\right)U_l
  \left(r\right)$, the radial equation turns out to be	
	\begin{equation}
		\frac{1}{\left(r^2+n^2\right)}\frac{d}{dr}\left[\left(r^2-2mr-n^2\right)
\frac{dU_l}{dr}\right]-\left[\frac{l\left(l+1\right)}{n^2+r^2}-\frac{\lambda^2}{4}Rgb^2\right]U_l=0 \,.
	\end{equation}
Here, we only consider the zero mode $l=0$. We require the probe scalar field is regular on the horizon and have the form
\begin{equation}
		U_l\left(r\right)=u_0+\frac{3u_0\left(n^2-r_h^2\right)\lambda^2}{r_h\left(n^2+r_h^2\right)^2} \left(r-r_h\right)+{\cal O} (r-r_h)^2 + \dots \,.
	\end{equation}
The probe scalar equation can not be solved analytically, but can be solved numerically. Given the value of $(n\,,\lambda)$, one can numerically obtain the value of $m$. We plot these bifurcation points in Fig.1, which shows the existence line of the scalarized black hole.
\begin{figure}[H]
		\centering
		\includegraphics[width=0.9\linewidth]{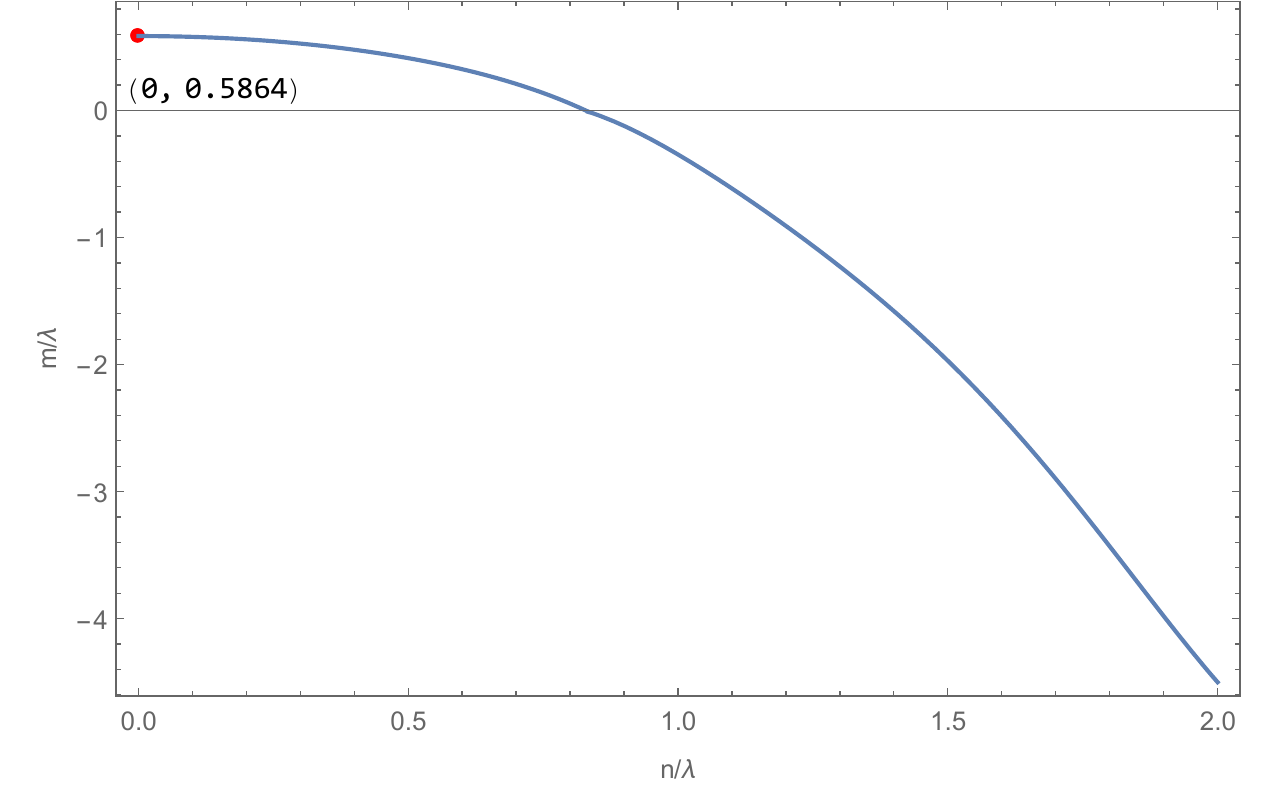}
		\caption{The line of bifurcation point where the new hairy black hole will start to emerge, the point of ($n=0$, $m/\lambda = 0.5864$) covers the result of Schwarzschild case. }
	\end{figure}
When $n=0$, $m/\lambda = 0.5864$, it is consistent with the Schwarzschild case\cite{Doneva:2017bvd}. Starting from these bifurcation points, we shall construct numerically the hairy black hole in the next section.

\section{Scalarization of Taub-NUT black hole}
 With the guidance in the previous section, we are now in a stage to obtain new hairy Taub-NUT black holes in extended scalar-tensor-Gauss-Bonnet theory. We consider the Taub-NUT like metric
	\begin{equation}
		ds^{2}=-f\left(r\right)\left(dt+2n\cos\theta d\phi\right)^{2}+\frac{dr^2}{h\left(r\right)}+\left(r^2+n^2\right)\left(d\theta^2+\sin^2\theta d\phi^2\right) \,,
	\end{equation}
with $h(r)$ and $f(r)$ are undetermined functions of $r$.  And the scalar field is static and spherical $\varphi = \varphi(r)$.
	Substituting these into the equations of motion, we can get four independent field equations	
	\newpage
	\begin{flalign}
		&\
		4fh^2r\left[2n^2\varPsi_r+r(n^2+r^2)\varPsi_{r}^{'}\right]-3n^2f^2(n^2+r^2)\left(-1+2\varPsi_rh^{'}+4h\varPsi_{r}^{'}\right)-&\nonumber
		\\
		&\ fh(n^2+r^2)^3\varphi^{'2}-fh(n^2+r^2)\left[2n^2+r^2-6r^2\varPsi_rh^{'}+4(n^2+r^2)\varPsi_{r}^{'}\right]+&\nonumber
		\\
		&\
		f(n^2+r^2)^2\left[1-\left(r+2\varPsi_r\right)h^{'}\right]=0,&
	\end{flalign}
	\begin{flalign}
		&\
		4f^2h\left[n^2+(n^2+r^2)^2\varphi^{'2}\right]-h(n^2+r^2)(n^2+r^2-4hr\varPsi_r)f^{'2}+2f^2r(n^2+r^2)h^{'}-&\nonumber
		\\
		&\
		4n^2f^3\left(2\varPsi_rh^{'}+4h\varPsi_r^{'}-1\right)+f(n^2+r^2)^2f^{'}h^{'}+2fhr(n^2+r^2)\left(1-6\varPsi_r^{'}h^{'}\right)f^{'}+&\nonumber
		\\
		&\
		2fh(n^2+r^2)^2f^{''}-8fh^2r(n^2+r^2)\varPsi_rf^{''}-8fh^2\left[n^2\varPsi_r+r(n^2+r^2)\varPsi_r^{'}\right]f^{'}=0,&
	\end{flalign}
	\begin{flalign}
		&\
		fh(n^2+r^2)^4\varphi^{'}f^{'}+f^2\left(n^2+r^2\right)^4\varphi^{'}h^{'}+2f^2h\left(n^2+r^2\right)^2\left[2r(n^2+r^2)\varphi^{'}+(n^2+r^2)^2\varphi^{''}\right]+&\nonumber
		\\
		&\
		\left\{fh(n^2+r^2)^2\left[3r^2\lambda^2h^{'}f^{'}-2\lambda^2(n^2+r^2)f^{''}\right]-3n^2\lambda^2ff^{'}\left(n^2+r^2\right)^2\left(f^{'}h+fh^{'}\right)+\right.&\nonumber
		\\
		&\
		\phantom{=\;\;}
		\left.4n^2\lambda^2f^3\left[2h\left(n^2-3r^2\right)+r\left(n^2+r^2\right)h^{'}\right]+\lambda^2h(n^2+r^2)(n^2+r^2-r^2h)f^{'}+\right.&\nonumber
		\\
		&\
		\phantom{=\;\;}
		\left.2\lambda^2fh^2r(n^2+r^2)\left[2n^2f^{'}+r(n^2+r^2)f^{''}\right]-\lambda^2f\left(n^2+r^2\right)^3f^{'}h^{'}+\right.&\nonumber
		\\
		&\
		\phantom{=\:\:}
		\left.24n^2rf^2h(n^2+r^2)^2f^{'}-6n^2\lambda^2f^2h\left(n^2+r^2\right)^2f^{''}\right\}\frac{dF(\varphi)}{d\varphi}=0,&
	\end{flalign}
	\begin{flalign}
		&\
		(n^2+r^2)h\left[r(n^2+r^2)+2(n^2+r^2-3r^2h)\varPsi_r\right]f^{'}-n^2f^2(n^2+r^2+8rh\varPsi_r)-(n^2+r^2)^2f-&\nonumber
		\\
		&\
		(n^2+r^2)fh\left[-r^2-6n^2f^{'}\varPsi_r+(n^2+r^2)^2\varphi^{'2}\right]=0,&
	\end{flalign}

	where $'$ is derivative with respect to radial coordinate $r$ and $\varPsi_r$ is defined by
	
	\begin{equation}
		\varPsi_r=\lambda^2\frac{dF(\varphi)}{d\varphi}\frac{d\varphi}{dr}  \,.
	\end{equation}

	It's difficult to obtain analytic solutions for field equations (15)-(18) with a nontrivial scalar field, thus, we will turn to numerical method and  use shooting method to construct new hairy black hole solutions. We shall explore the behavior of the fields on both boundaries.

  On the horizon $r_h$, we  require the metric functions $h$ and $f$ vanish simultaneously and impose regularity conditions for $\varphi$
  	\begin{equation}
		f\left(r\right)|_{r\rightarrow\\r_h}\rightarrow 0,\quad
		h\left(r\right)|_{r\rightarrow\\r_h}\rightarrow 0,\quad
		\varphi\left(r\right)|_{r\rightarrow\\rh}\rightarrow \varphi_h  \,.
	\end{equation}
	Taking Taylor expansion near the event horizon $r_h$, the equations of motion tell us that the fields have the form
	 \begin{align}
		&\
		h\left(r\right)=\frac{1+\delta}{r_h} \left(r-r_h\right)+{\cal O}(r-r_h)^2  \,, \cr
		&f\left(r\right)=\frac{r_h\left[e^{\left(-12\varphi_h^2\right)}\left(n^2+r_h^2\right)\delta
+6\varphi_h^2\left(1+\delta\right)^2\lambda^4\right]}{6n^2 \varphi_h^2\left(1+\delta\right)\lambda^4}\left(r-r_h\right)+{\cal O}(r-r_h)^2   \,, \cr
		&\varphi\left(r\right)=\varphi_h-\frac{e^{6\varphi_h^2}r_h\delta}{2\varphi_h\left(1+\delta\right)
\lambda^2}\left(r-r_h\right)+{\cal O}(r-r_h)^2 \,. \label{exh}
	\end{align}
There are three parameters on the horizon, $r_h$, $\delta$ and $\varphi_h$. It is worth pointing out that  the effect of the Gauss-Bonnet coupling term appears in the leading term of the fields.

   In the infinity, we take a large $r$ expansion and let the fields be	
	\begin{align}
		&\
		h\left(r\right)=h_0+\frac{h_1}{r}+\frac{h_2}{r^2}+\frac{h_3}{r^3}+\frac{h_4}{r^4}+\frac{h_5}{r^5}+...
		\\
		&f\left(r\right)=f_0+\frac{f_1}{r}+\frac{f_2}{r^2}+\frac{f_3}{r^3}+\frac{f_4}{r^4}+\frac{f_5}{r^5}+...
		\\
		&\varphi\left(r\right)=\varphi_0+\frac{
\varphi_1}{r}+\frac{\varphi_2}{r^2}+\frac{\varphi_3}{r^3}+\frac{\varphi_4}{r^4}+\frac{\varphi_5}{r^5}+...
	\end{align}
	
	Substituting into the field equations (15-18), and solving the equations of motion order by order in $1/r$, actually, we calculate up to order $1/r^6$. It is turns out that $h_1=f_1$ and the coefficients ($h_i\,, f_i\,,\varphi_i$) can be expressed in terms of ($f_1\,,\varphi_1$) for $i\ge2$. Usually, $f_1$ is related to black hole mass while $\varphi_1$ is related to scalar charge, thus we rename them to be $f_1=-2m$ and $\varphi_1=D$. We show the coefficients to fifth order below
	\begin{align}
		&\
		h\left(r\right)=1-\frac{2M}{r}+\frac{D^2-2n^2}{r^2}+\frac{M\left(D^2+2n^2\right)}{r^3}+\frac{\left(4D^2 M^2+6n^4+2n^2D^2\right)}{3r^4}\nonumber
		\\
		&\quad\quad\quad+\frac{M}{12r^5}\left\{63D^4-4D^2\left(142M^2+65n^2-4\lambda^2\right)+8\left(80M^4+12n^2-51n^2\lambda^2\right)\nonumber\right.
		\phantom{=\;\;}
		\\
		&\left.\quad\quad\quad+8M^2\left(80n^2-21\lambda^2\right)\right\}+...,
		\\
		&f\left(r\right)=1-\frac{2M}{r}-\frac{2n^2}{r^2}+\frac{M\left(D^2+6n^2\right)}{3r^3}+\frac{2D^2M^2+2n^2\left(D^2+3n^2\right)}{3r^4}\nonumber
		\\
		&\quad\quad\quad+\frac{M}{60r^5}\left\{55D^4-4D^2\left(130M^2+50n^2-28\lambda^2\right)+8\left(80M^4-51n^2\lambda^2\right)\nonumber\right.
		\phantom{=\;\;}
		\\
		&\left.\quad\quad\quad+8M^2\left(80n^2-2\lambda^2\right)\right\}+...,
		\\
		&\varphi\left(r\right)=\frac{D}{r}+\frac{DM}{r^2}-\frac{D\left[D^2-2\left(4M^2+n^2\right)\right]}{6r^3}+\frac{DM\left(6M^2-2D^2+3n^2\right)}{3r^4}\nonumber
		\\
		&\quad\quad\quad+\frac{D}{120r^5}\left\{9D^4+24\left[16M^4+n^4+3n^2\lambda^2+3M^2\left(4n^2-\lambda^2\right)\right]\nonumber\right.
		\phantom{=\;\;}
		\\
		&\left.\quad\quad\quad-4D^2\left(58M^2+13n^2\right)\right\}+...
	\end{align}
Here, we can see that the coupling constant $\lambda$ appears in order $1/r^5$, which implies the Gauss-Bonnet term has minor effect in the infinity.

We are now in a stage to construct new black holes numerically, we shall apply numerical method to connect the horizon data to the ones in the infinity. One choice is to integrate the solution from horizon to infinity. We can perform a Taylor expansions on the horizon and use it as initial data, then integrate the solution from it to the infinity. However, one can see from the complexity of $h_1$ and $f_1$, we can not push the Taylor expansion to much higher orders to guarantee the accuracy. Thus, we take a alternative direction of integrating the solution from infinity to the middle, since the series expansion in the infinity can be carried to a much higher order (We actually calculate up to order of $1/r^6$, and higher than this order is reachable.). The emergence of a black hole solution is signified by the simultaneously vanishing of $f(r)$ and $h(r)$.

\section{Numerical Results}
	
With the guidance of the probe scalar field analysis under the Taub-NUT solution, we start to search new Taub-NUT black hole around the bifurcation point and do find a new branch Taub-NUT like black hole solutions with scalar hair. In Fig.2, we show the metric functions $f\left(r\right)$, $h\left(r\right)$ and scalar field $\varphi\left(r\right)$ for positive and negative mass parameter respectively. It is obvious that the metric functions have a peak higher than 1 for negative mass parameter, whilst there is no peak for positive mass parameter. In Fig.3, we show the metric function $f$ for fixed NUT parameter $n=1$ and various value of mass parameter. As can one see, a peak emerges on the profile of metric function from bottom to top as the mass decreases  from a positive value to a negative value. The event horizon radius becomes smaller as the mass parameter decreases, which is under expectation.	
	\begin{figure}[H]
		\centering
		\subfloat[$m>0$]{
			\centering
			\includegraphics[width=0.5\linewidth]{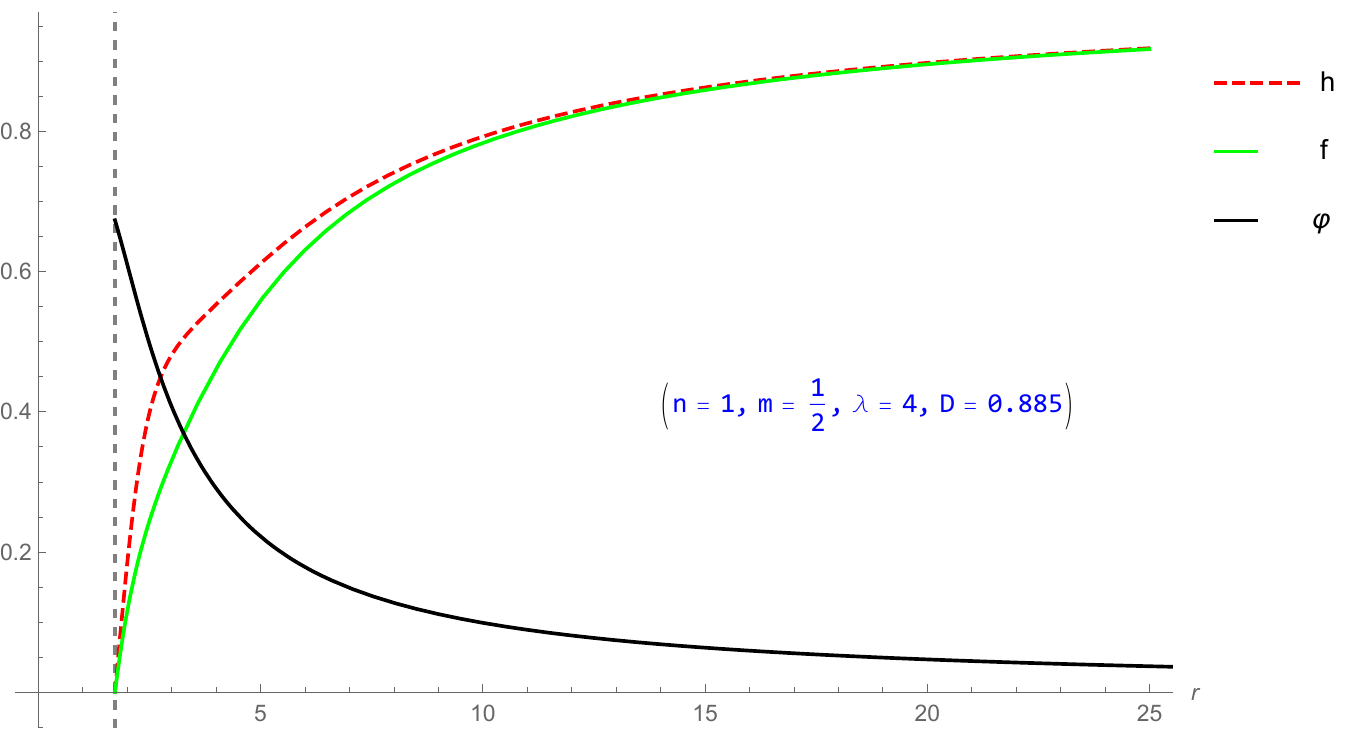}	
		}
		\subfloat[$m<0$]{
			\includegraphics[width=0.5\linewidth]{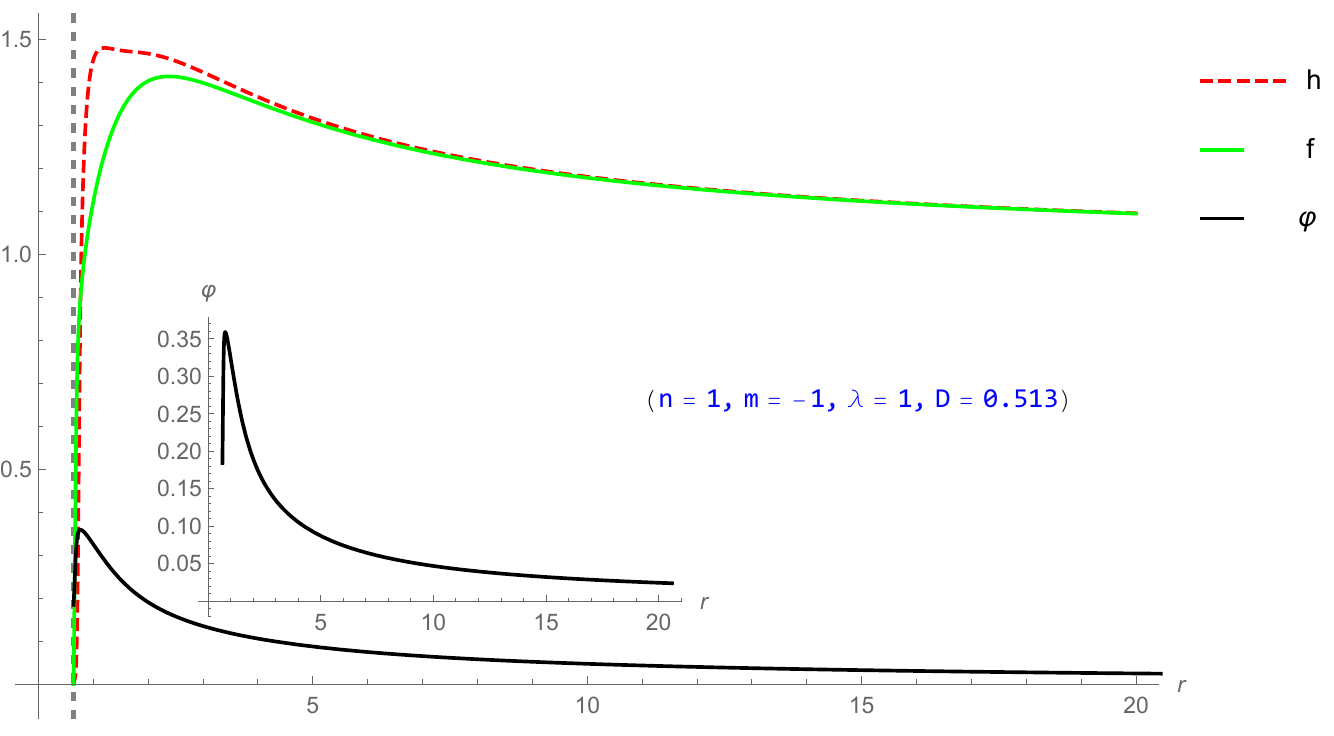}
		}
		\caption{The newly constructed Taub-NUT black hole solutions for two different type of mass parameters. The left represents $m>0$, the right one is for $m<0$, the profile (h,f) of which have a peak. }
	\end{figure}
		\begin{figure}[H]
		\centering
		\includegraphics[width=0.8\linewidth]{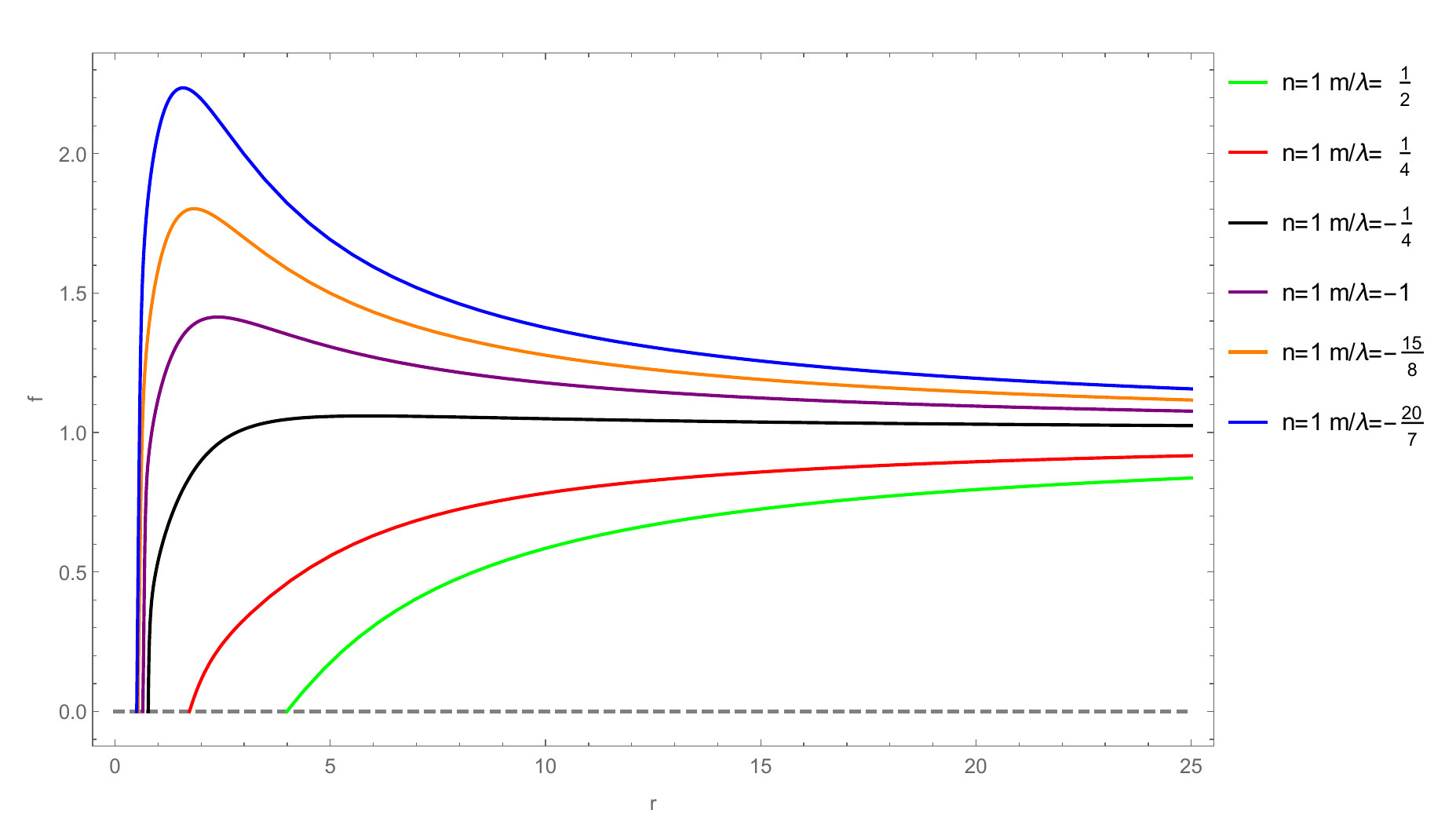}
		\caption{The metric function $f\left(r\right)$ for fixed NUT parameter n and various mass parameter $m$, the one with negative mass parameter has a peak. }
	\end{figure}

We plot the mass parameter $m$ as a function of black hole radius $r_h$ in Fig.4. Since the theory admits analytic Taub-NUT black hole as a solution, we also include the $m-r_h$ curve of Taub-NUT black hole as a comparison. The solid line represents Taub-NUT black hole, while the dashed line is the new scalarized black hole. The new black holes start from the bifurcation point which is consistent with the probe scalar analysis. There exists two particular properties for our new constructed black holes compared with the Schwarzschild case. One is that there is a minimal value of $r_h$ for the new black holes. Another is there are two branches of new solutions and these two new black holes smoothly connected to each other at the minimal radius. The situation changes when we increase the value of NUT parameter $n$. There is just one branch of new solution for large $n$, however, there still exists a minimal value of black hole radius. We present the $m-r_h$ curve for various NUT parameters in the bottom of Fig.4, including the $n=0$ case.

	\begin{figure}[H]
		\centering
		\subfloat[]{
			\centering
			\includegraphics[width=0.5\linewidth]{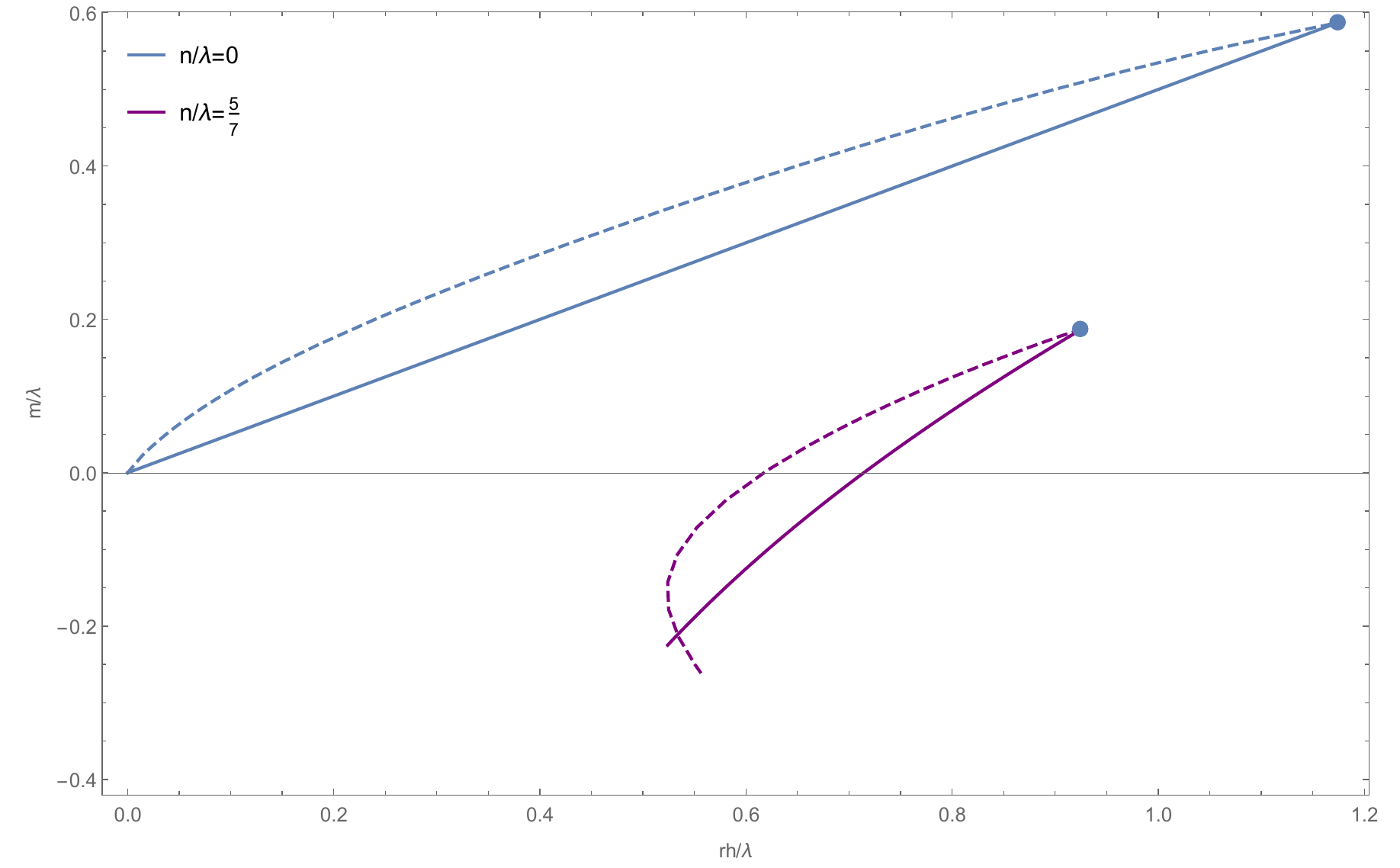}	
		}
		\subfloat[]{
			\includegraphics[width=0.5\linewidth]{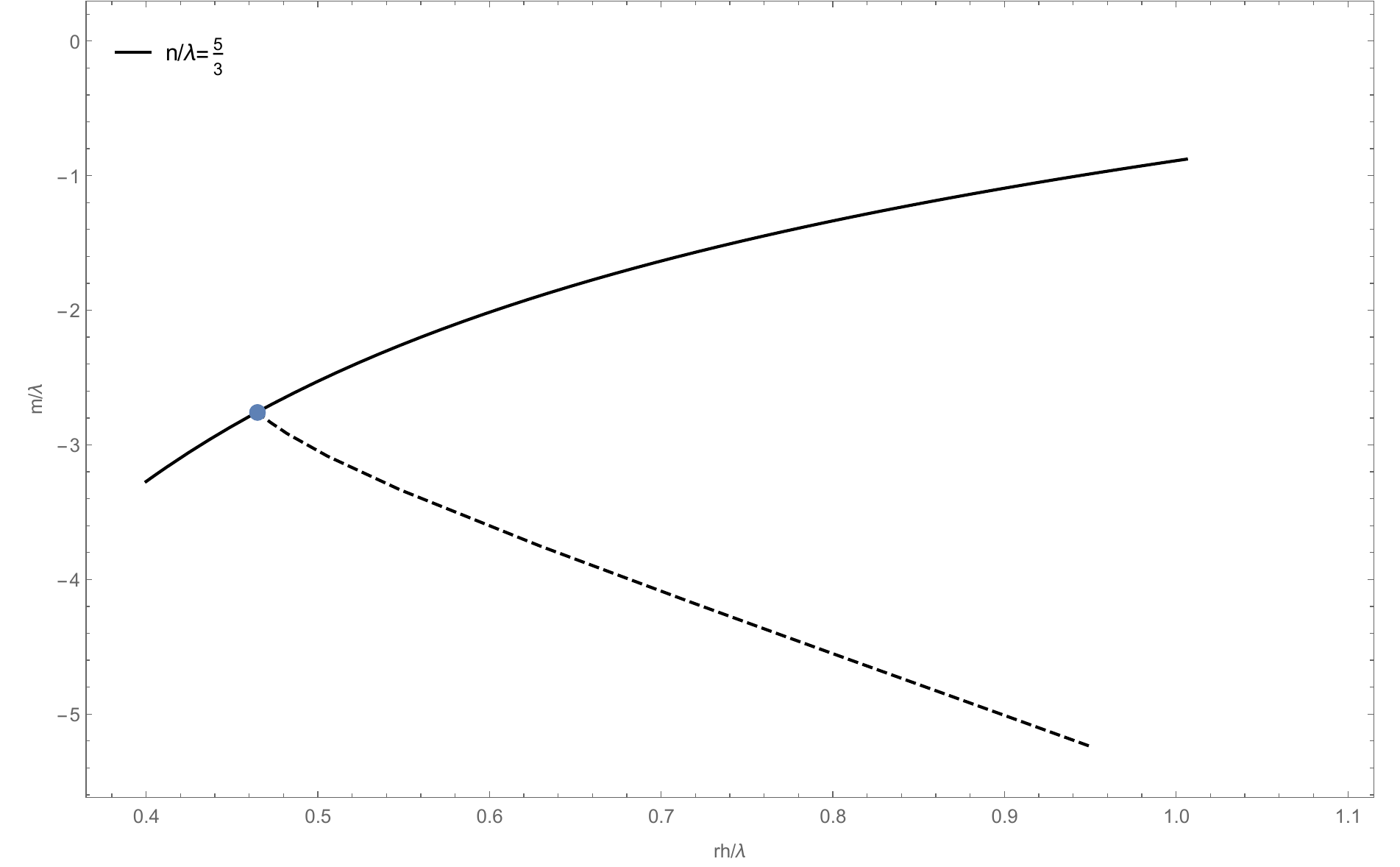}
		}
		\\
		\subfloat[]{
			\centering
			\includegraphics[width=0.8\linewidth]{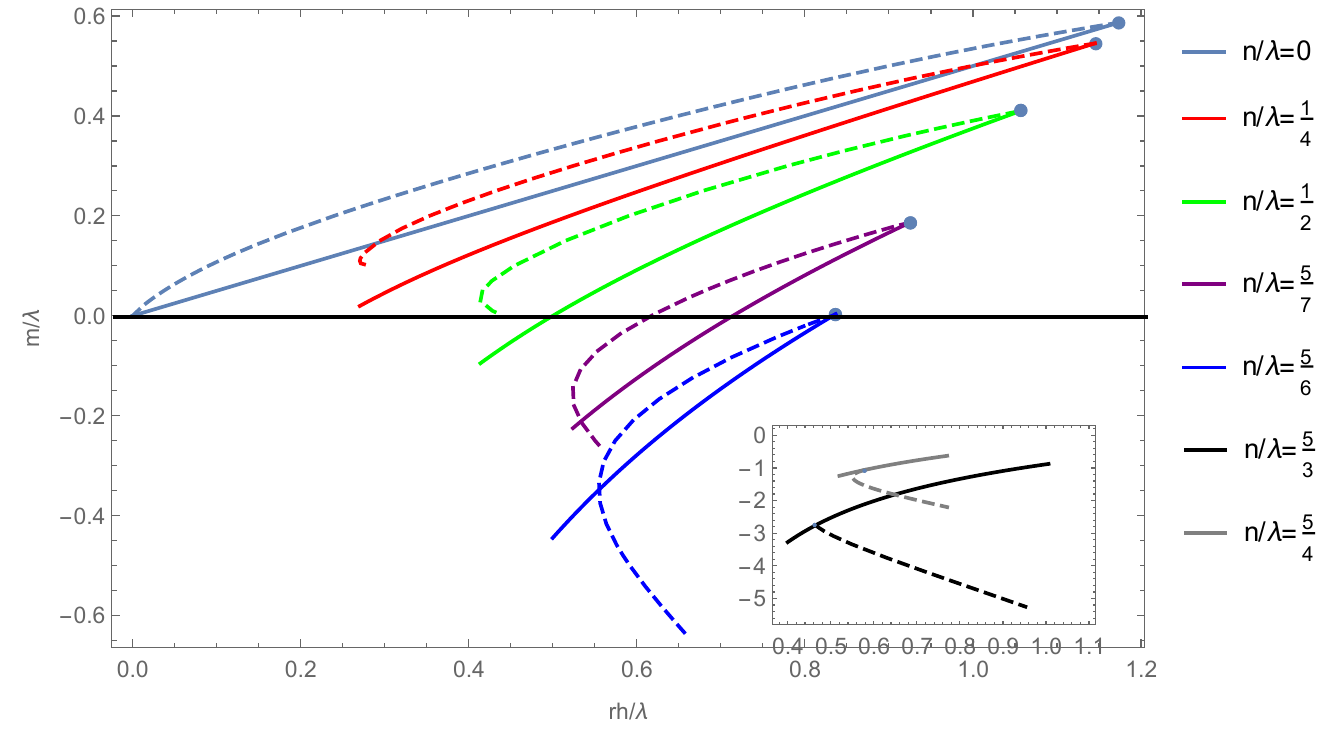}	
		}
		\caption{The mass parameter $m$ as a function of the horizon radius $r_h$ for different values of $n/\lambda$. (a) shows ($m-r_h$) curves for $n=0, and n/\lambda = 5/7$, the $n=0$ is for comparison. The dashed line represents the scalarized black hole, whilst the solid line represents the scalar-free black hole. The scalarized black hole for $n/\lambda = 5/7$ has two branches of solution, while the Schwarzschild case of $n=0$ has only one branch. (b) shows the curve for $n/\lambda = 5/3$, which has only one branch new solution, the notation is the same, the dashed line represents the scalarized black hole, the other is for scalar-free black hole.   (c) shows  ($m-r_h$) curves for various NUT parameter. }
	\end{figure}

Next, we want to explore the properties of the scalar hair of the new black hole. It is obvious that the theory is invariant under $\varphi \rightarrow - \varphi$, thus the scalar charge parameter $D$ is symmetric.  We first plot the scalar charge parameter $D$ as a function of mass parameter $m$ for various NUT parameter in the top of Fig.5.  We also plot the case for $n=0$, which is closed, the scalar charge $D$ increases from zero to maximal value and then decreases back to zero as mass parameter decreases.  Whilst, the curve is not closed for our new found black holes. The scalar charge $D$ of our new black hole increases from zero at the bifurcation point then terminates at certain value of $m$. From these curves, we can see that every scalarized black hole has a maximal value of mass parameter, which is at the bifurcation point. From this $D-m$ curve, we can not distinguish the two branches of the new black hole solution, since they are smoothly connected. Then, we plot the scalar charge parameter $D$ as a function of black hole radius $r_h$ in the bottom of Fig.5, where we can see clearly that there exists two branches of new black hole solutions with a minimal $r_h$ where these two are connected. One can see that the second branch is short and not obvious when $n$ is small (n=0 corresponding to Schwarzschild case, where there is only one branch of hairy black hole solution.), and the second branch is more obvious when $n$ is large. However, there exists only one branch black hole solution when n is large enough.
	
	\begin{figure}[H]
		\centering
		\subfloat[]{
			\includegraphics[width=0.5\linewidth]{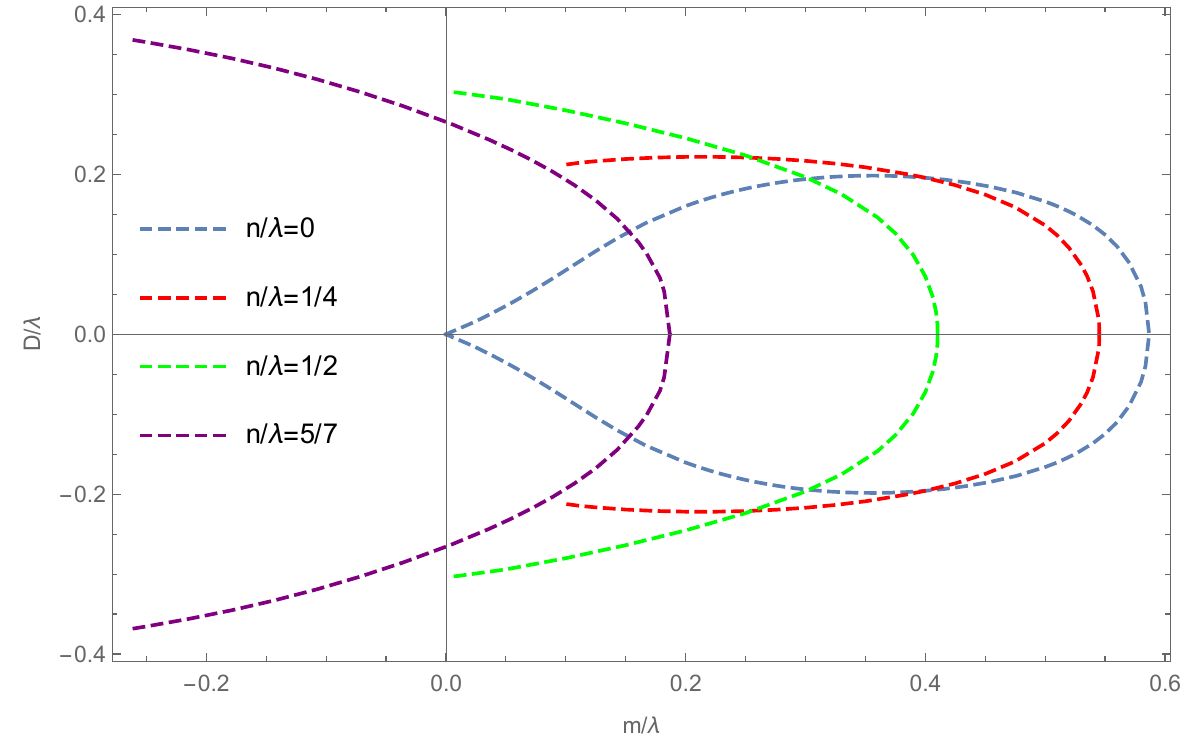}
		}
		\subfloat[]{
			\includegraphics[width=0.5\linewidth]{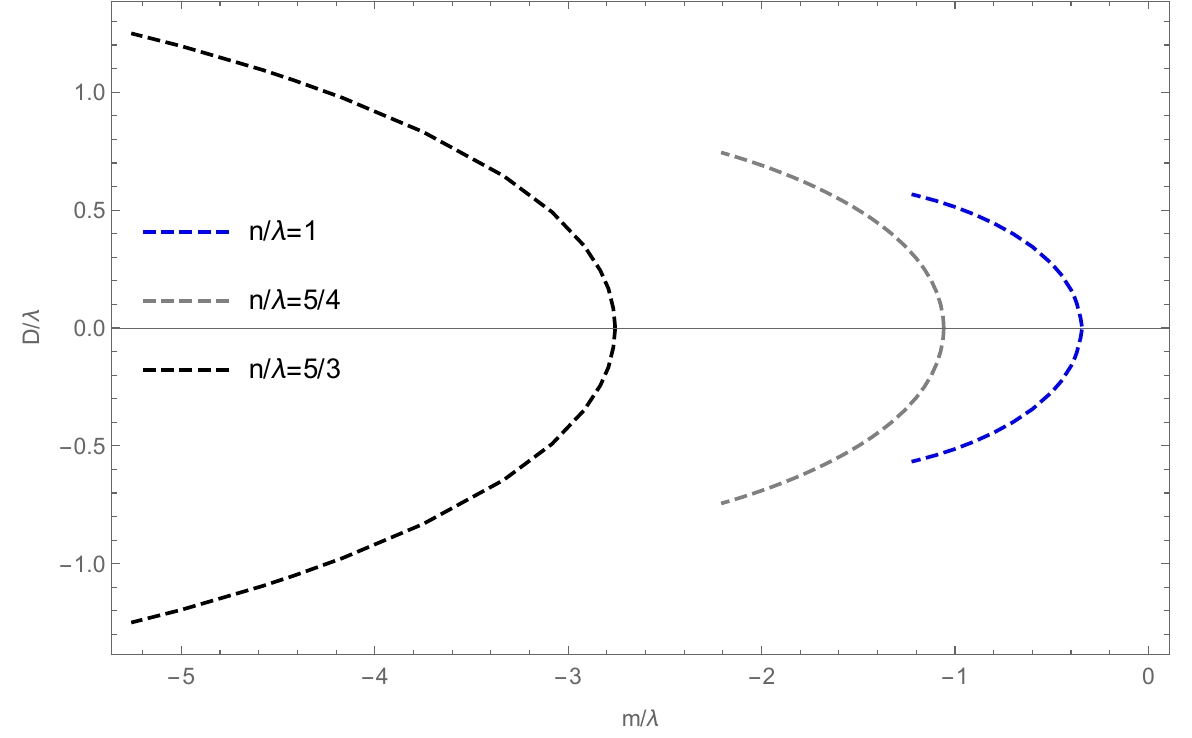}
		}
		\\
		\subfloat[]{
			\includegraphics[width=0.5\linewidth]{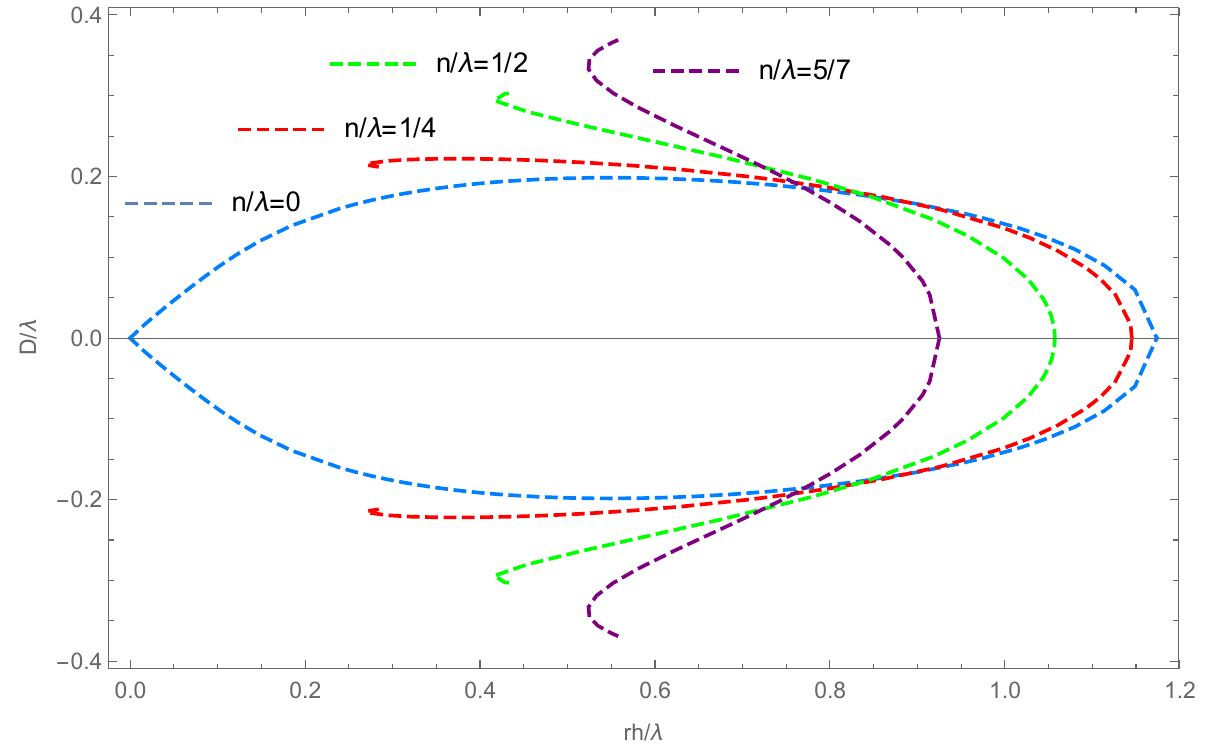}
		}
		\subfloat[]{
			\includegraphics[width=0.5\linewidth]{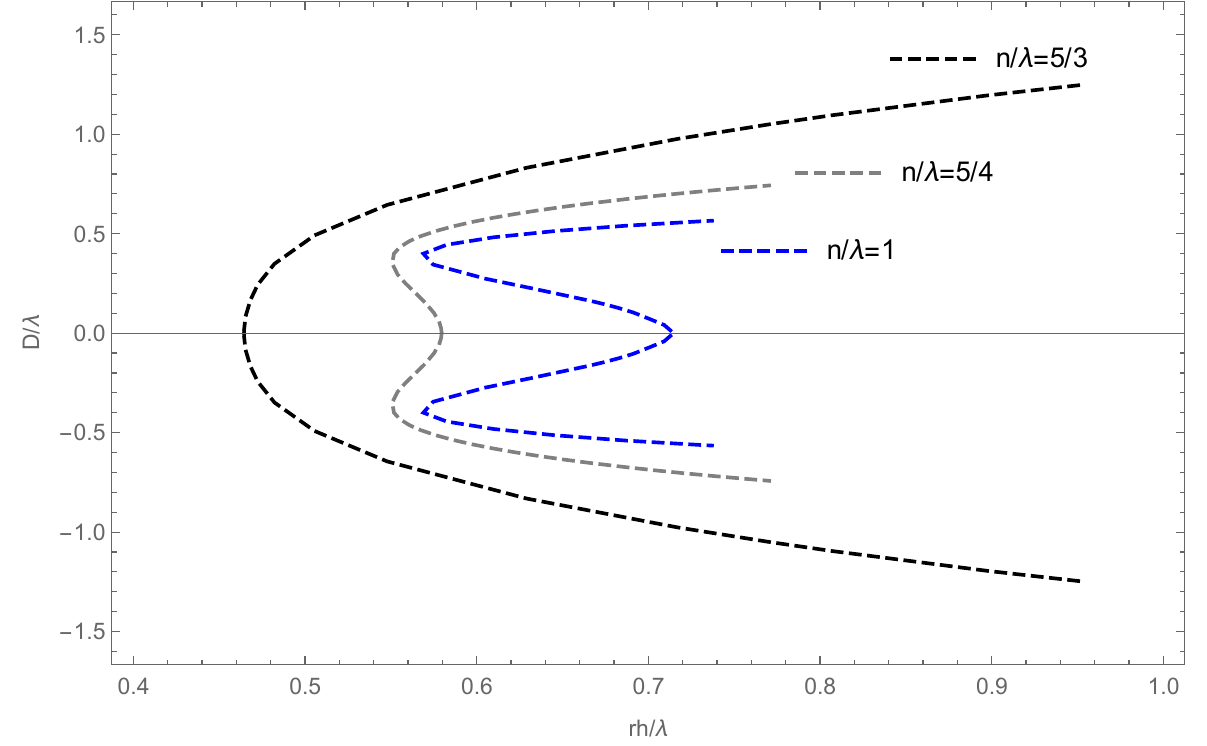}
		}
		\caption{The scalar charge $D$ as functions of the mass parameter $m$  is shown in the top two figures, all the scalarized black hole have a maximal value of mass parameter.  The scalar charge $D$ as functions of the horizon radius $r_h$ for different values of $n/\lambda$ in bottom two figures,  one can see the scalarized black holes have two branches which of smoothly connected to each other for small value of n, while have only one branch when the NUT parameter is large.}
	\end{figure}

After exploring the basic properties of the new black holes, we turn to the thermodynamic properties of the solutions, the entropy and temperature. Since the presence of higher curvature Gauss-Bonnet term in the action, the entropy is no longer equal to one quarter of the event horizon area $A_h$ anymore. The correct entropy of black hole can be calculated through Wald entropy formula\cite{Wald:1993nt,Iyer:1994ys}
\begin{equation}
		S_H=2\pi\int_{r_h}\left(r^2+n^2\right)^2 \epsilon_{ab}\epsilon_{cd} \frac{\partial L}{\partial R_{abcd}} \,,
	\end{equation}
where L is the Lagrangian and $\epsilon_{ab}$ is the volume form on the 2-dimensional cross section $r_h$ of the horizon. For the extended scalar-tensor-Gauss-Bonnet theory we consider, the explicit expression of black hole entropy is given by
\begin{equation}
		S_H=\pi(r_h^2+n^2)+4\pi\lambda^2F\left(\varphi_h\right) \,.
	\end{equation}
It is obvious that the entropy is not equal to one quarter of event horizon area in general, and reduces back to one quarter of event horizon when the Guass-Bonnet term is turned off by setting $\lambda =0$.

Before presenting result for entropy, we want to show the relations between the event horizon area $A_h$ and the mass parameter $m$ of the new found black holes in Fig.6. As one can see, the event horizon areas of the scalarized black hole (dashed lines) are smaller than that of scalar-free black hole(solid lines) for small value of NUT parameter $n$. As the NUT parameter $n$ is increasing, the situation changes, and finally the areas of scalarized black holes are large than that of scalar-free black holes. We notice another novel phenomena that  once the mass parameter $m$ is positive $(m>0)$ at the bifurcation points where the scalarized black holes begin, the areas of event horizon are almost the same for different NUT parameters at the bifurcation points, see the horizontal line in Fig.6. 	
		\begin{figure}[H]
		\centering
		\includegraphics[width=0.8\linewidth]{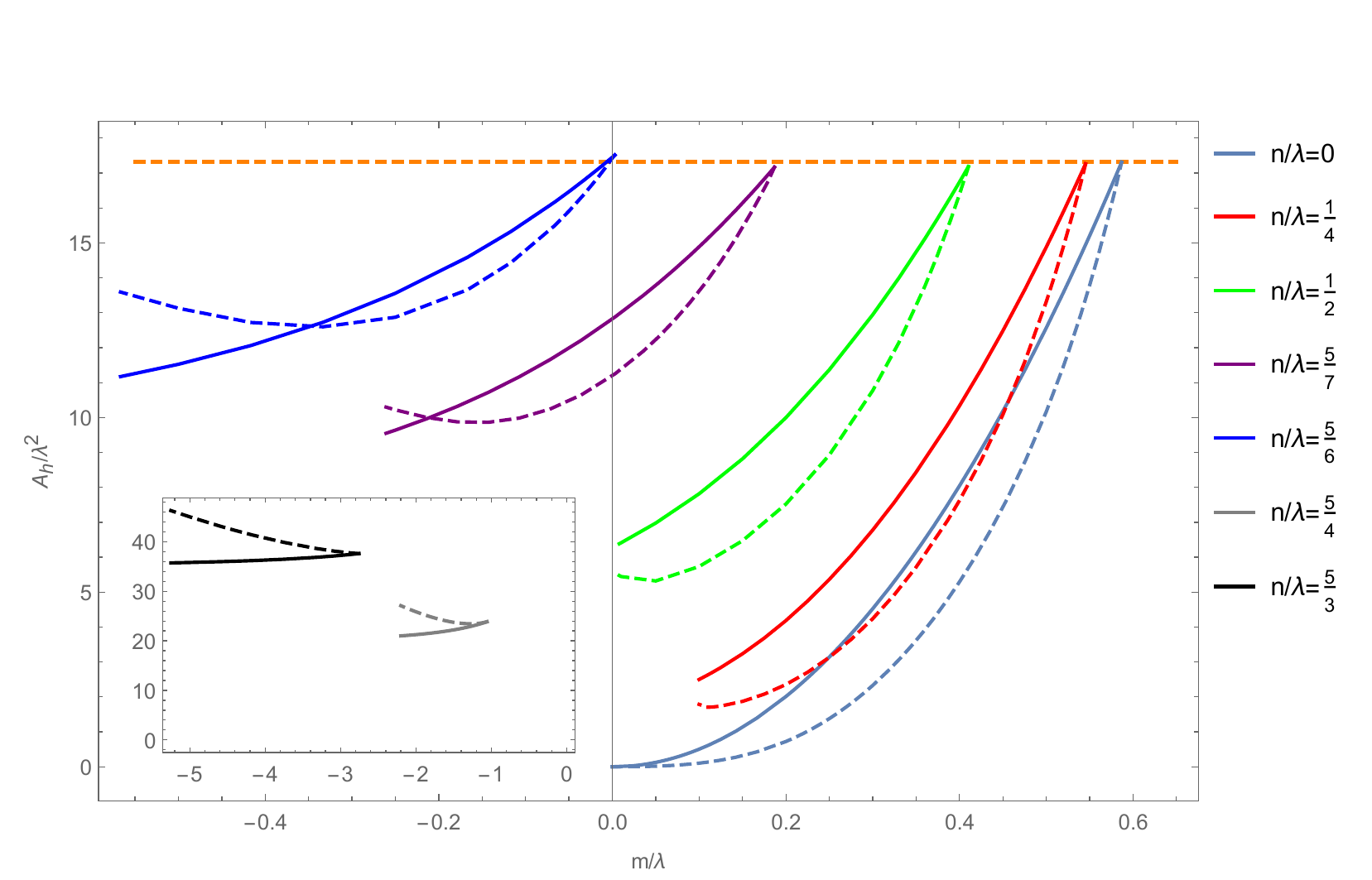}
		\caption{The area of the black hole horizon $A_H$ as functions of the mass parameter $m$ for different values of $n/\lambda$, the value of area is almost the same at the starting point for positive mass parameter. }
	\end{figure}
At this stage, we present the results of entropy in Fig.7. We plot entropy $S$ as a function of mass parameter $m$ in the left panel of Fig.7. There is one big difference between the area-mass relation $A_h-m$ in Fig.6. Here, the entropy of scalarized black hole is always larger than that of scalar-free black holes, the reason is that the Gauss-Bonnet term has an additional contribution to entropy for scalarized black holes. The hairy black holes with larger entropy implies that they are more thermodynamically stable and more favorable than the scalar-free black holes. The right panel of Fig.7 shows the entropy as a function of event horizon radius. From the right panel, we can see two branches of new hairy black hole solutions smoothly connected, again. At the bifurcation points, the similar phenomenon emerges, the entropy at bifurcation points are nearly the same, see the horizontal line in Fig.7.. The value of entropy at bifurcation point is about 4.32 for positive mass parameter, and it is  also the largest entropy for the scalarized black hole. It is mentioned in the beginning of this section that every scalarized black hole for fixed NUT parameter has a maximal value of mass parameter which is at the bifurcation point. We conjecture that all the saclarized black holes with the maximal mass parameter is greater than zero have a maximal entropy bound of 4.32 which is located at the bifurcation point.

	\begin{figure}[H]
		\centering
		\subfloat[$S/\lambda^2$ as a function of $m/\lambda$]{
			\centering
			\includegraphics[width=0.5\linewidth]{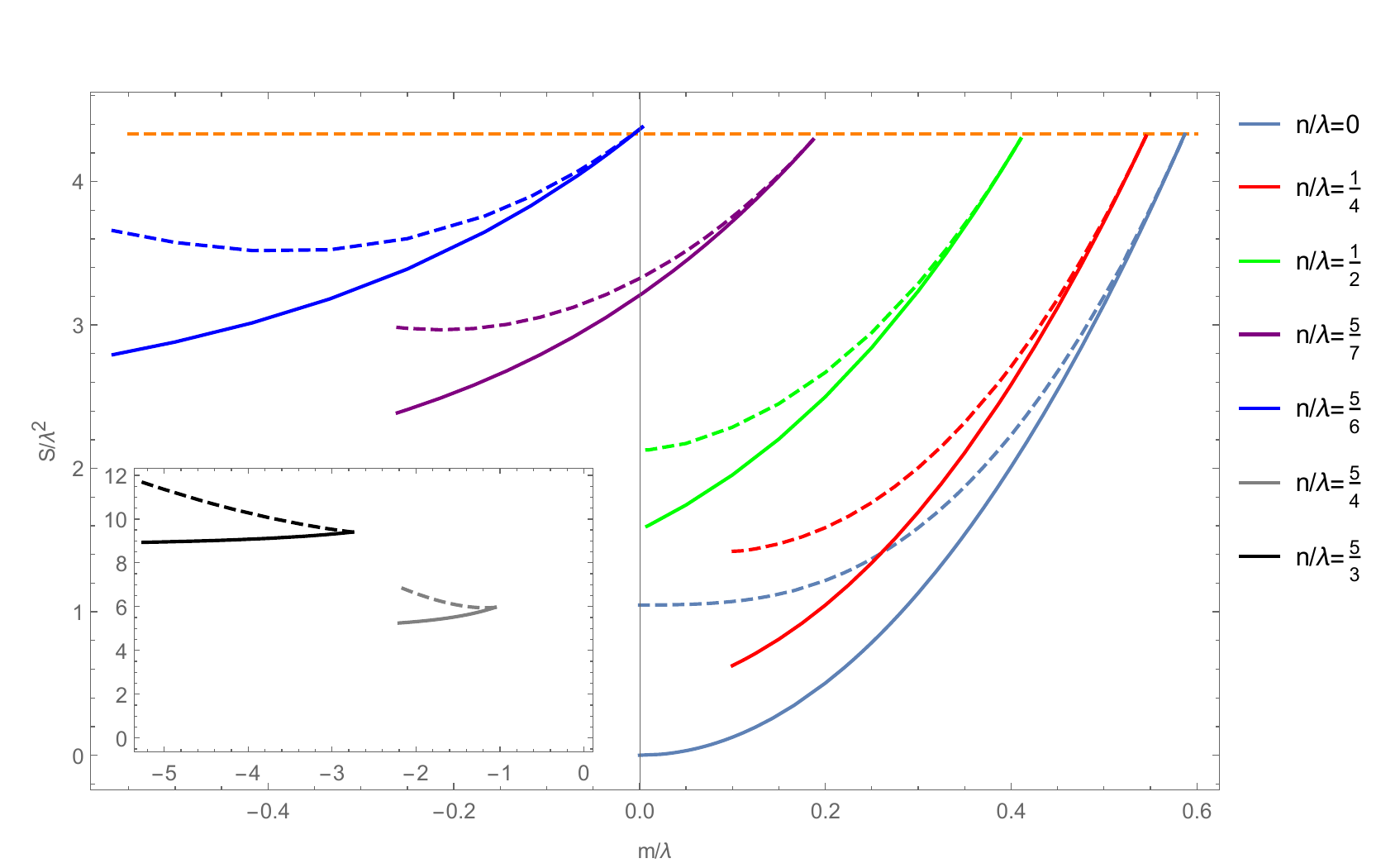}	
		}
		\subfloat[$S/\lambda^2$ as a function of $r_h/\lambda$]{
			\includegraphics[width=0.5\linewidth]{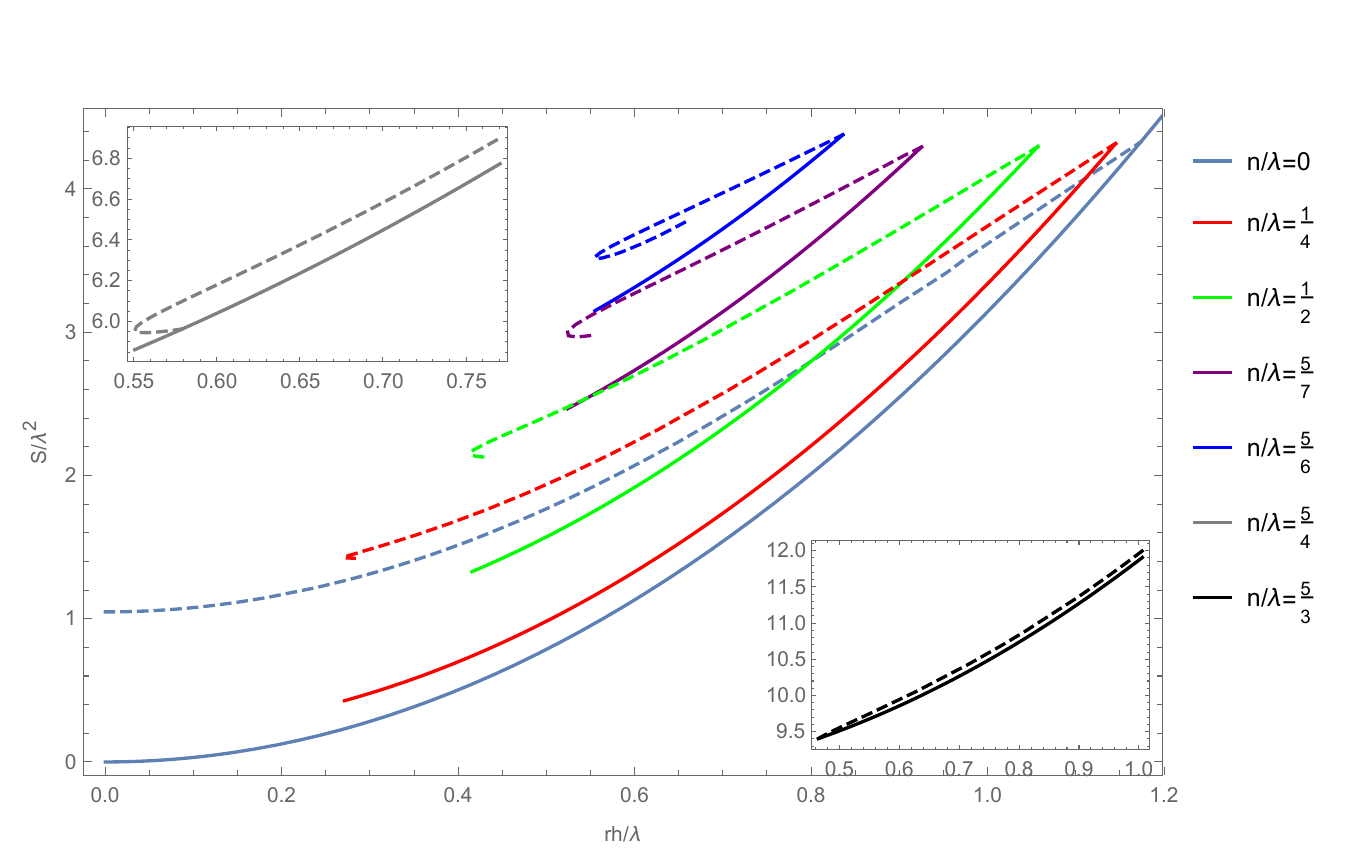}
		}
		\caption{The entropy $S$ of the black hole as functions of the mass parameter $m$ is shown in the left figure, the entropy $S$ as functions of the horizon radius $r_h$ is shown in the right figure. The entropy is nearly a constant at the bifurcation point for positive mass parameter.}
	\end{figure}

The temperature can be calculated through standard method, for our Taub-NUT like metric, it is given by
\begin{eqnarray}
		T = \frac{\sqrt{h'(r_h) f'(r_h)}}{4\pi} \,.
	\end{eqnarray}	
The temperature is plotted as a function of mass parameter and event horizon radius respectively in Fig.8.  As is seen, the temperature of the hairy black holes is always higher than that of scala-free black holes. In the right panel, we can see that the temperature of one branch hairy black holes is always higher than that of the other branch. 	
	\begin{figure}[H]
		\centering
		\subfloat[$\lambda T$ as a function of $m/\lambda$]{
			\centering
			\includegraphics[width=0.5\linewidth]{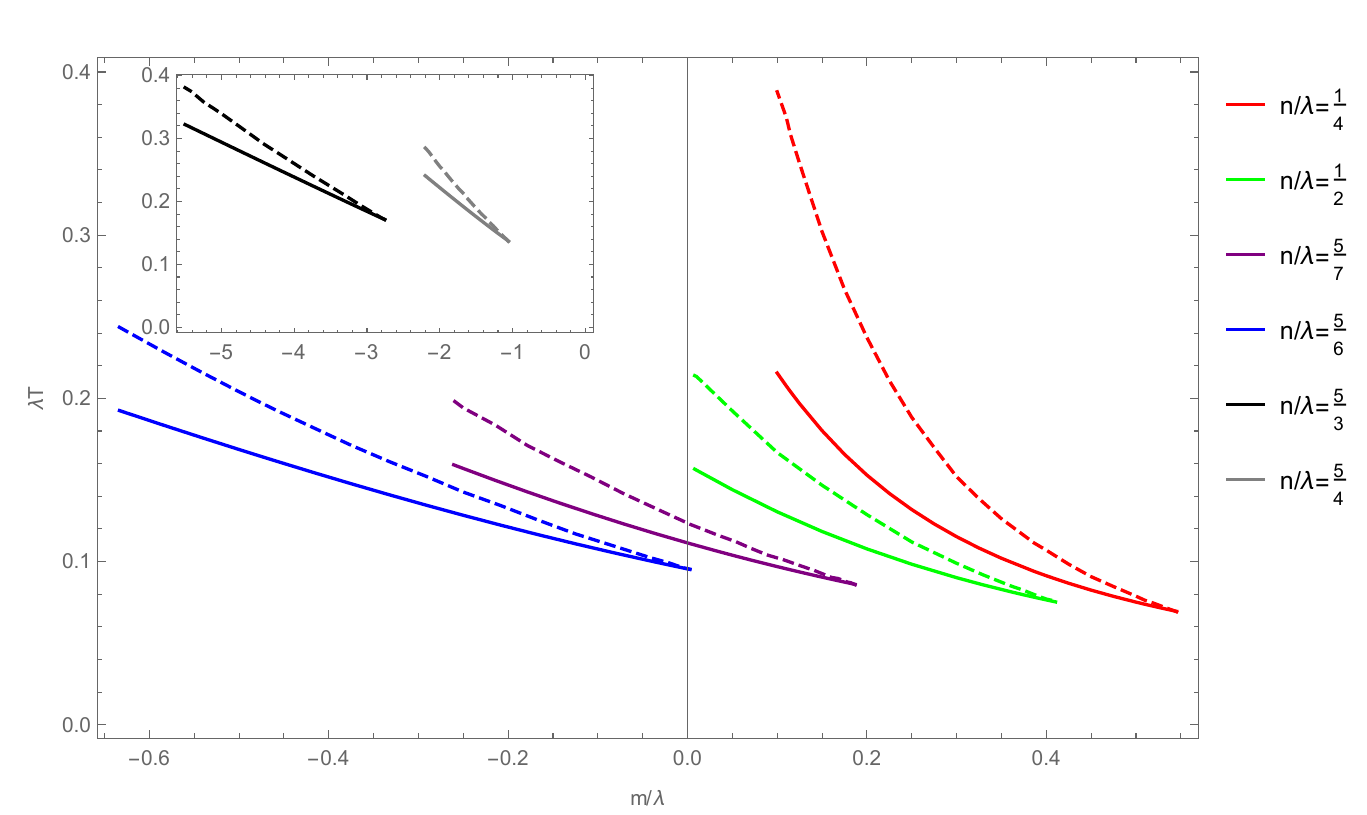}	
		}
		\subfloat[$\lambda T$ as a function of $r_h/\lambda$]{
			\includegraphics[width=0.5\linewidth]{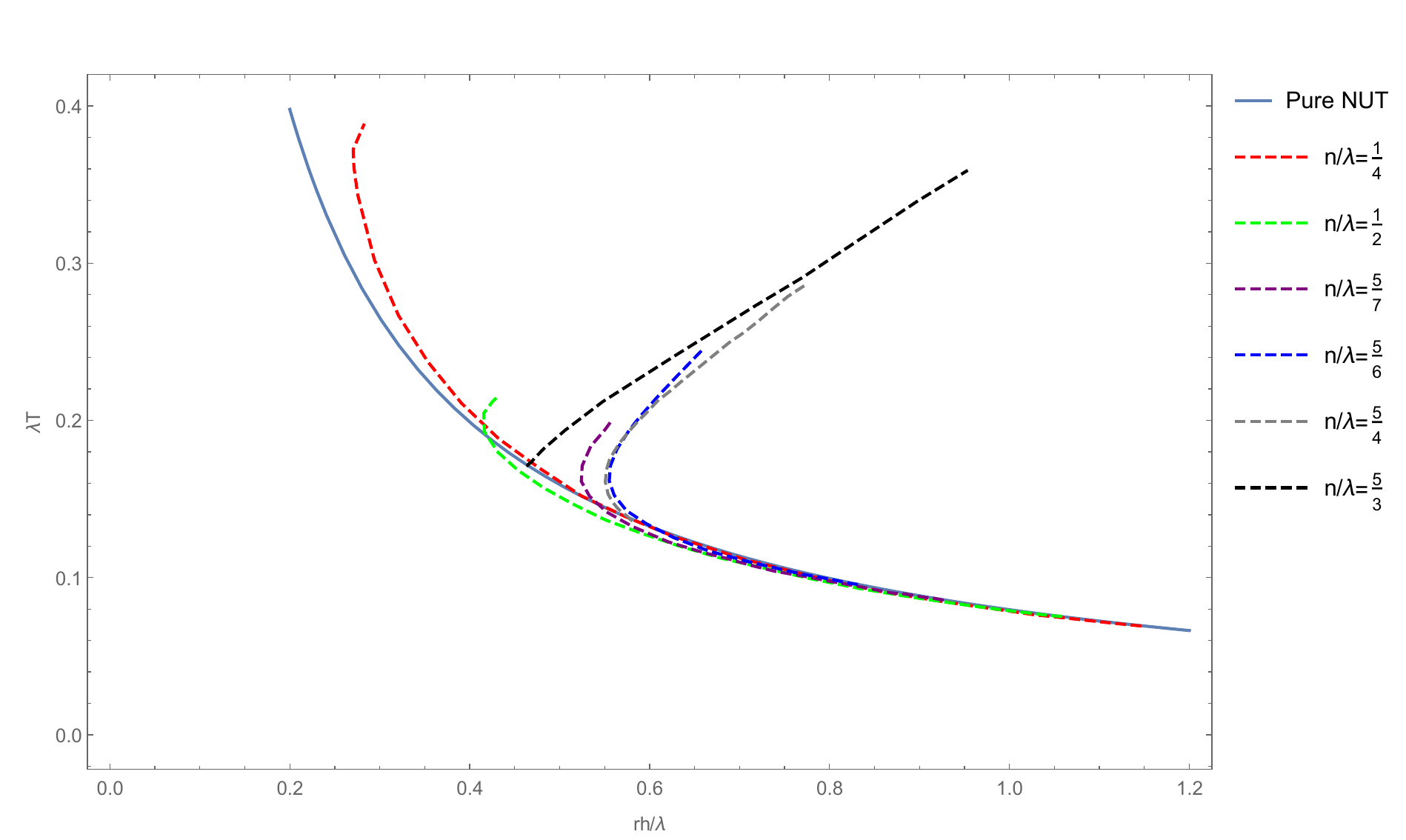}
		}
		\caption{The temperature $T$ of the black hole as functions of the mass parameter $m$ and the horizon radius $r_h$ for different values of $n/\lambda$.}
	\end{figure}
Through the probe scalar analysis, we can find the bifurcation points of hairy black holes and scalar-free black holes where the scalarized black holes begin, connecting these bifurcation points gives the existence line. From all the results we collect, the hairy black holes terminate at some points,  these points constitute the critical line. We plot these two line in Fig.9, between these two lines is the "habitat" of the new scalarized black hole solution. 	
\begin{figure}[H]
		\centering
		\includegraphics[width=0.8\linewidth]{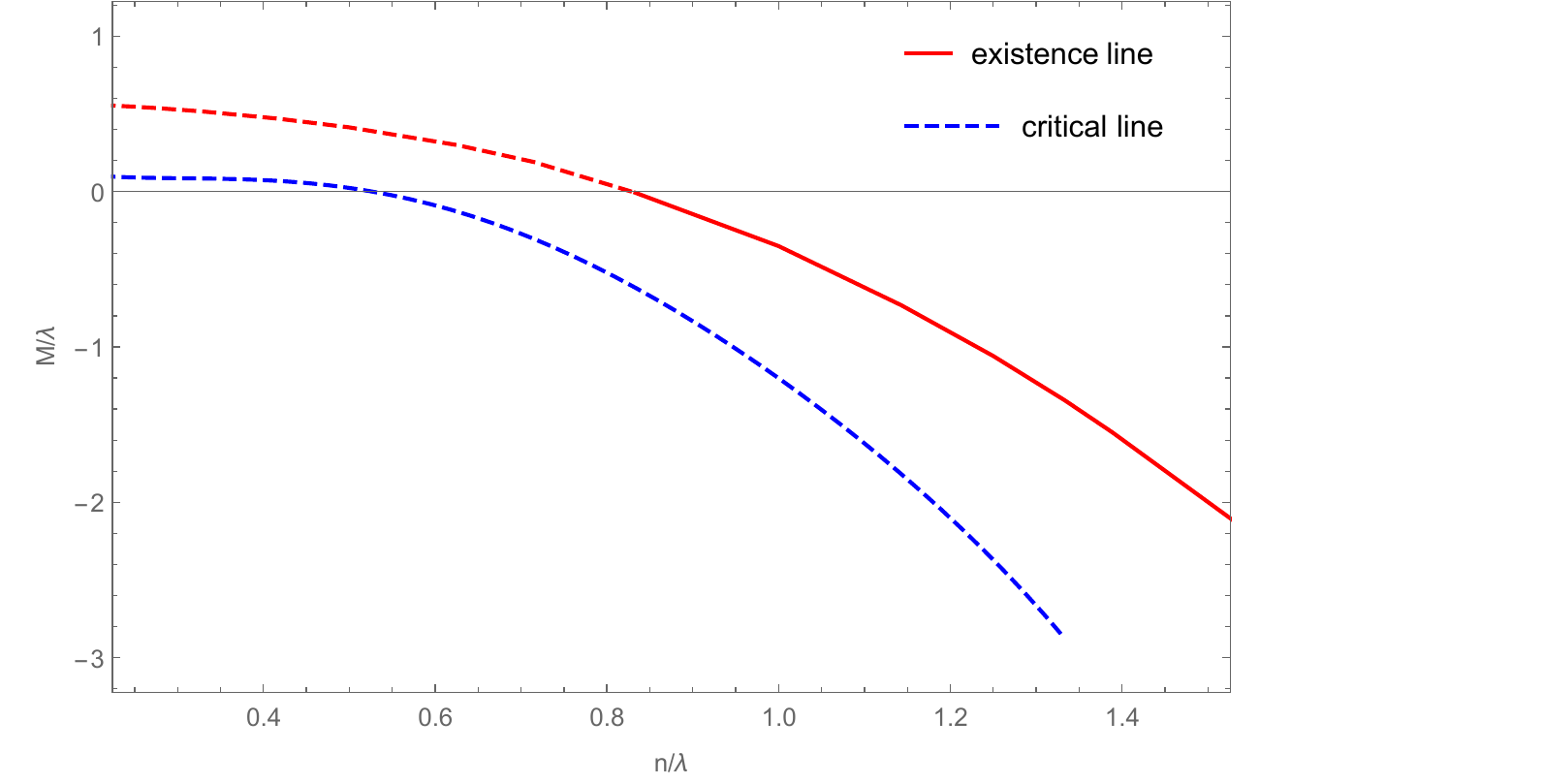}
		\caption{The scalarized Taub-NUT blac holes start from the red line, the value of mass parameter is positive for the dashed red line, while the solid red line represents the part with negative mass parameter.  The hairy Taub-NUT black holes terminate at the blue line. }
	\end{figure}
	As we pointed out in the entropy part, the entropy and horizon area of black hole in the bifurcation point are nealy constant for positive mass parameter, which means $r_h^2+n^2$ remains constant. Since it happens in the bifurcation points, we can trade $r_h$ with mass parameter $m$ using the relation $\fft{r_h^2 - 2 m r_h - n^2}{r_h^2+n^2}=0$ on the event horizon, then we get the relation between $m$ and $n$
	\begin{equation}
		\left(\frac{m}{\lambda}\right)^2+\left(\frac{n}{\lambda}\right)^2+
		\frac{m}{\lambda}\sqrt{\left(\frac{m}{\lambda}\right)^2+\left(\frac{n}{\lambda}\right)^2}\approx\frac{2.16}{\pi} \,.
	\end{equation}

	We plot the curve of this analytic relation together with the existence line in the left of Fig.10. One can see that the two curves almost overlap for positive mass, however, it deviates from each other when the mass parameter becomes negative. We also plot the difference of NUT parameters of the the two curves as a function of mass parameter in the right of Fig.10, the difference is obvious for negative mass parameter, whilst it drops rapidly to zero as mass parameter turns to positive value.

	\begin{figure}[H]
		\centering
		\subfloat[]{
			\centering
			\includegraphics[width=0.5\linewidth]{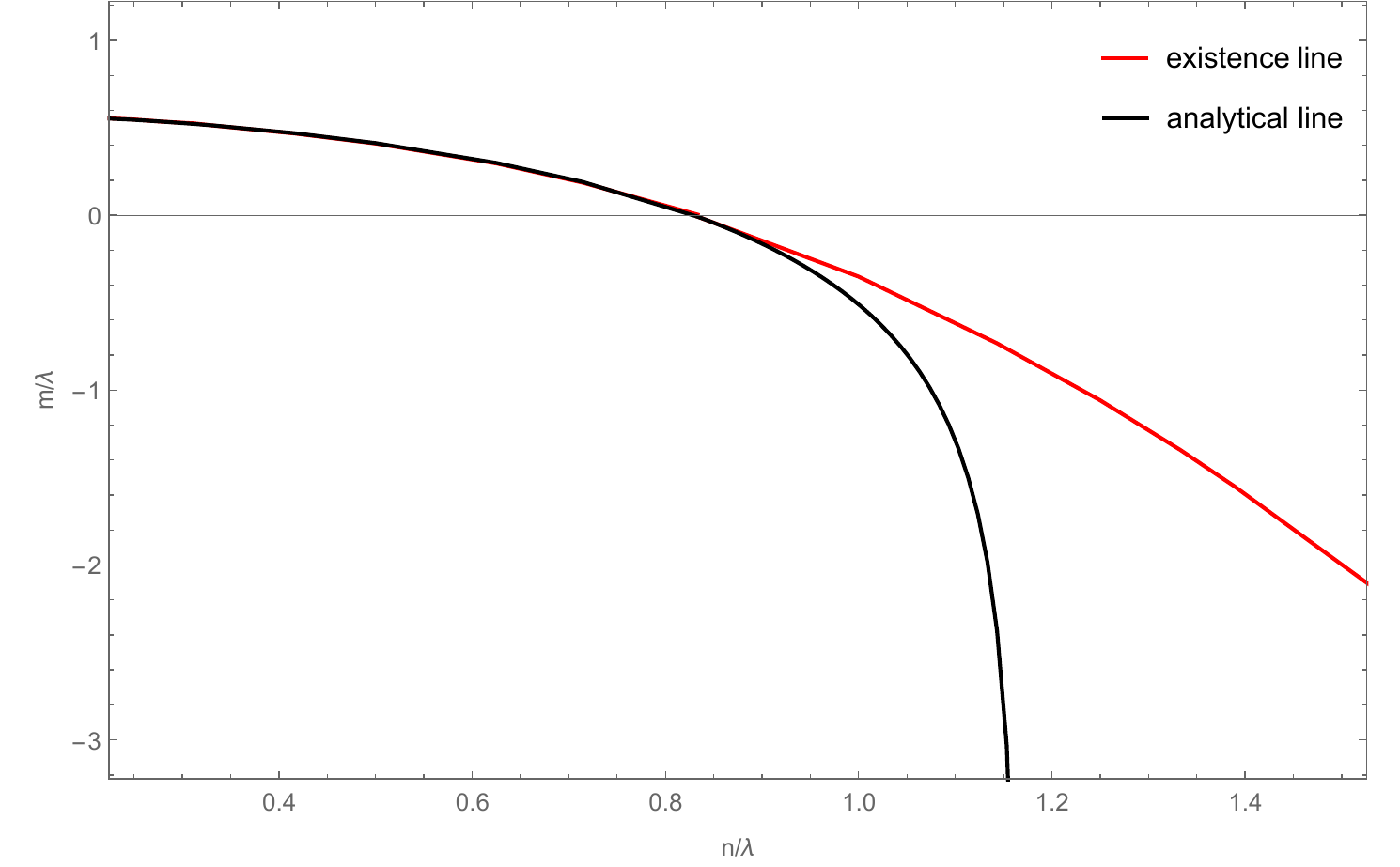}	
		}
		\subfloat[]{
			\includegraphics[width=0.49\linewidth]{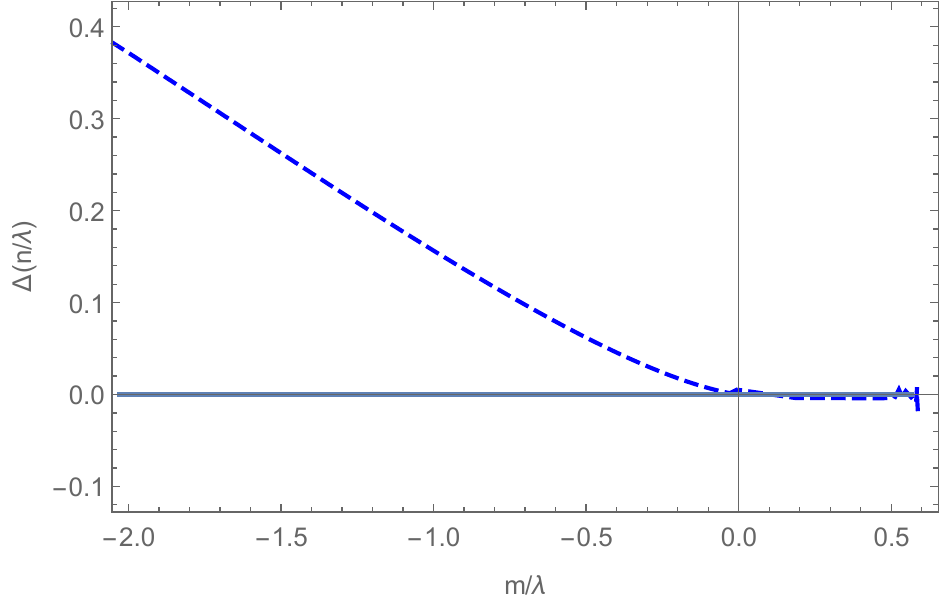}
		}
		\caption{The left panel shows the bifurcation points where the scalarized black hole begin and the curve for constant entroy, they lie on top of each other for positive mass parameter. The right panel shows difference of NUT parameter of existence line and constant entropy line, it is almost zero for positive mass parameter. }
	\end{figure}

	\section{conclusions}
In this work we have studied the scalarization of Taub-NUT black holes in the extended scalar-tensor-Gauss-Bonnet gravity theory, where the nontrivial scalar is excited by the curvature of the spacetime through the Gauss-Bonnet coupling term. We choose a coupling function such that the theory admits the Ricci-flat Taub-NUT black hole as a solution.  Guided by the probe scalar analysis, we construct the hairy Taub-NUT black holes numerically. We observe one novel phenomenon that there exists two branches of new hairy black hole solutions which are smoothly connected to each other under certain parameter range. As far as we know, there is no such phenomenon appears in the scalarization of Schwarzschild black holes. The NUT parameter $n$ plays a significant role in this phenomenon. As far as the NUT parameter is not zero,  the second branch of the hairy black hole emerges, whilst there is no trace of the second branch as NUT parameter vanishes.  When the NUT parameter is small, the second branch hairy black hole is not obvious, and it becomes more obvious as NUT parameter increases. However, when the NUT parameter is large enough, the second branch solution disappears, again.

The obtained hairy Taub-NUT black holes bifurcate from the analytic Taub-NUT black holes, but do not fall back to the scalar-free Taub-NUT black holes like the scalarization of Schwarzschild. Instead, they terminate at some point for all the cases we have studied. The bifurcation point is shifted to smaller mass parameter as the NUT parameter increases.

We have also  studied the entropy and temperature of the hairy black holes. The temperature of scalarized black holes is always larger than scalar-free Taub-NUT black holes.  We found that the event horizon area of hairy black holes is not always larger than that of scalar-free Taub-NUT black holes, but entropy is, the reason is that the Gauss-Bonnet term has a additional contribution to the entropy of hairy black holes. This implies that the new found black hole solutions are thermodynamically more  stable. We also observe a novel phenomena in this system that the entropy of the black holes at the bifurcation point is almost constant for positive mass parameter. We then conjecture a maximal entropy bound for all the scalarized black hole whose mass parameter at the bifurcation point is greater than zero. The physics under this phenomena is worth exploring further.

The definition of  mass and NUT charge of Taub-NUT black holes in Einstein's general relativity is still controversial, many progress has been made in this direction\cite{Liu:2022wku,Hennigar:2019ive,Wu:2019pzr,Chen:2019uhp}. Turning to the higher curvature theories, such as the theory we consider, the situation is even more complicated. The new scalarized black holes supply one more platform to explore the definition of mass and NUT charge in higher derivative gravity theories.

\section{Acknowledgement}

We are grateful to Yu-Qi Chen, Ze Li, Jun-Fei Liu and Hong Lu for useful discussion. This work is supported in part by NSFC (National Natural Science Foundation of China) Grant No.~12075166.


\begin{thebibliography}{99}
	\bibitem{Bekenstein:1972ny}
	J.~D.~Bekenstein,
	\textit{Transcendence of the law of baryon-number conservation in black hole physics},
	Phys. Rev. Lett. \textbf{28}, 452-455 (1972)
	
	\bibitem{Bekenstein:1995un}
	J.~D.~Bekenstein,
	\textit{Novel \textquoteleft{}\textquoteleft{}no-scalar-hair\textquoteright{}\textquoteright{} theorem for black holes},
	Phys. Rev. D \textbf{51}, no.12, R6608 (1995)

	\bibitem{Doneva:2017bvd}
	D.~D.~Doneva and S.~S.~Yazadjiev,
	\textit{New Gauss-Bonnet Black Holes with Curvature-Induced Scalarization in Extended Scalar-Tensor Theories},
	Phys. Rev. Lett. \textbf{120}, no.13, 131103 (2018)
	[arXiv:1711.01187 [gr-qc]].
	
	\bibitem{Antoniou:2017hxj}
	G.~Antoniou, A.~Bakopoulos and P.~Kanti,
	\textit{Black-Hole Solutions with Scalar Hair in Einstein-Scalar-Gauss-Bonnet Theories},
	Phys. Rev. D \textbf{97}, no.8, 084037 (2018)
	[arXiv:1711.07431 [hep-th]].
	
	\bibitem{Silva:2017uqg}
	H.~O.~Silva, J.~Sakstein, L.~Gualtieri, T.~P.~Sotiriou and E.~Berti,
	\textit{Spontaneous scalarization of black holes and compact stars from a Gauss-Bonnet coupling},
	Phys. Rev. Lett. \textbf{120}, no.13, 131104 (2018)
	[arXiv:1711.02080 [gr-qc]].
	
	\bibitem{Minamitsuji:2018xde}
	M.~Minamitsuji and T.~Ikeda,
	\textit{Scalarized black holes in the presence of the coupling to Gauss-Bonnet gravity},
	Phys. Rev. D \textbf{99}, no.4, 044017 (2019)
	[arXiv:1812.03551 [gr-qc]].
	
	\bibitem{Cunha:2019dwb}
	P.~Cunha, V.P., C.~A.~R.~Herdeiro and E.~Radu,
	\textit{Spontaneously Scalarized Kerr Black Holes in Extended Scalar-Tensor\textendash{}Gauss-Bonnet Gravity},
	Phys. Rev. Lett. \textbf{123}, no.1, 011101 (2019)
	[arXiv:1904.09997 [gr-qc]].
	
	\bibitem{Myung:2018vug}
	Y.~S.~Myung and D.~C.~Zou,
	\textit{Instability of Reissner\textendash{}Nordstr\"om black hole in Einstein-Maxwell-scalar theory},
	Eur. Phys. J. C \textbf{79}, no.3, 273 (2019)
	[arXiv:1808.02609 [gr-qc]].
	
	\bibitem{Damour:1993hw}
	T.~Damour and G.~Esposito-Farese,
	\textit{Nonperturbative strong field effects in tensor - scalar theories of gravitation},
	Phys. Rev. Lett. \textbf{70}, 2220-2223 (1993)
	
	\bibitem{Taub:1950ez}
	A.~H.~Taub,
	\textit{Empty space-times admitting a three parameter group of motions},
	Annals Math. \textbf{53}, 472-490 (1951)
	
	\bibitem{Newman:1963yy}
	E.~Newman, L.~Tamburino and T.~Unti,
	\textit{Empty space generalization of the Schwarzschild metric},
	J. Math. Phys. \textbf{4}, 915 (1963)

\bibitem{NUTthmann1}
R.~A.~Hennigar, D.~Kubiz\v{n}\'ak and R.~B.~Mann,``Thermodynamics of Lorentzian Taub-NUT spacetimes,''Phys. Rev. D \textbf{100}, no.6, 064055 (2019)
doi:10.1103/PhysRevD.100.064055
[arXiv:1903.08668 [hep-th]].

\bibitem{NUTth2}
A.~Ballon Bordo, F.~Gray, R.~A.~Hennigar and D.~Kubiz\v{n}\'ak,``The First Law for Rotating NUTs,''Phys. Lett. B \textbf{798}, 134972 (2019)
doi:10.1016/j.physletb.2019.134972
[arXiv:1905.06350 [hep-th]].


\bibitem{NUTth3}
A.~Awad and S.~Eissa,
``Lorentzian Taub-NUT spacetimes: Misner string charges and the first law,''
Phys. Rev. D \textbf{105}, no.12, 124034 (2022)
doi:10.1103/PhysRevD.105.124034
[arXiv:2206.09124 [hep-th]].

	
	\bibitem{Liu:2022wku}
	H.~S.~Liu, H.~Lu and L.~Ma,
	\textit{Thermodynamics of Taub-NUT and Plebanski solutions},
	JHEP \textbf{10}, 174 (2022)
	[arXiv:2208.05494 [gr-qc]].
	
	\bibitem{Brihaye:2018bgc}
	Y.~Brihaye, C.~Herdeiro and E.~Radu,
	\textit{The scalarised Schwarzschild-NUT spacetime},
	Phys. Lett. B \textbf{788}, 295-301 (2019)
	[arXiv:1810.09560 [gr-qc]].
	
	\bibitem{Herdeiro:2018wub}
	C.~A.~R.~Herdeiro, E.~Radu, N.~Sanchis-Gual and J.~A.~Font,
	\textit{Spontaneous Scalarization of Charged Black Holes},
	Phys. Rev. Lett. \textbf{121}, no.10, 101102 (2018)
	[arXiv:1806.05190 [gr-qc]].
	
	\bibitem{Wald:1993nt}
	R.~M.~Wald,
	\textit{Black hole entropy is the Noether charge},
	Phys. Rev. D \textbf{48}, no.8, R3427-R3431 (1993)
	[arXiv:gr-qc/9307038 [gr-qc]].
	
	\bibitem{Iyer:1994ys}
	V.~Iyer and R.~M.~Wald,
	\textit{Some properties of Noether charge and a proposal for dynamical black hole entropy},
	Phys. Rev. D \textbf{50}, 846-864 (1994)
	[arXiv:gr-qc/9403028 [gr-qc]].

	\bibitem{Hennigar:2019ive}
	R.~A.~Hennigar, D.~Kubiz\v{n}\'ak and R.~B.~Mann,
	\textit{Thermodynamics of Lorentzian Taub-NUT spacetimes},
	Phys. Rev. D \textbf{100}, no.6, 064055 (2019)
	[arXiv:1903.08668 [hep-th]].

	\bibitem{Wu:2019pzr}
	S.~Q.~Wu and D.~Wu,
	\textit{Thermodynamical hairs of the four-dimensional Taub-Newman-Unti-Tamburino spacetimes},
	Phys. Rev. D \textbf{100}, no.10, 101501 (2019)
	[arXiv:1909.07776 [hep-th]].

	\bibitem{Chen:2019uhp}
	Z.~Chen and J.~Jiang,
	\textit{General Smarr relation and first law of a NUT dyonic black hole},
	Phys. Rev. D \textbf{100}, no.10, 104016 (2019)
	[arXiv:1910.10107 [hep-th]].
	
\end{thebibliography}
\end{document}